\documentclass[fleqn,usenatbib]{mnras}

\usepackage[T1]{fontenc}
\DeclareRobustCommand{\VAN}[3]{#2}
\let\VANthebibliography\thebibliography
\def\thebibliography{\DeclareRobustCommand{\VAN}[3]{##3}\VANthebibliography}

\usepackage{graphicx}	
\usepackage{amsmath}	
\usepackage{amssymb}	
\usepackage{xspace} 
\usepackage{xcolor}
\usepackage{upgreek}
\usepackage{CJK}
\usepackage{fontawesome}
\usepackage{gensymb}
\usepackage{multirow}
\usepackage{mathtools}
\usepackage{newtxtext,newtxmath}

\newcommand{\Gaia}{\textit{Gaia}\xspace} 

\title[Accreted stars in NIHAO and GALAH]{Finding accreted stars in the Milky Way: clues from NIHAO simulations}

\author[Buder, Mijnarends, and Buck]{
S. Buder,$^{1,2}$\thanks{E-mail: sven.buder@anu.edu.au}
L. Mijnarends,$^{1,2}$ and
T. Buck,$^{3,4}$
\\
$^{1}$Research School of Astronomy \& Astrophysics, Australian National University, ACT 2611, Australia\\
$^{2}$Center of Excellence for Astrophysics in Three Dimensions (ASTRO-3D), Australia\\
$^{3}$Universit{\"a}t Heidelberg, Interdisziplin{\"a}res Zentrum f{\"u}r Wissenschaftliches Rechnen, Im Neuenheimer Feld 205, D-69120 Heidelberg, Germany\\
$^{4}$Universit{\"a}t Heidelberg, Zentrum f{\"u}r Astronomie, Institut f{\"u}r Theoretische Astrophysik, Albert-Ueberle-Straße 2, D-69120 Heidelberg, Germany
}

\date{Accepted DD MM 2024. Received 22 04 2024; in original form 22 04 2024}

\pubyear{2024}

\begin{document}
\label{firstpage}
\pagerange{\pageref{firstpage}--\pageref{lastpage}}
\maketitle

\begin{abstract} 
Exploring the marks left by galactic accretion in the Milky Way helps us understand how our Galaxy was formed. However, finding and studying accreted stars and the galaxies they came from has been challenging. This study uses a simulation from the NIHAO project, which now includes a wider range of chemical compositions, to find better ways to spot these accreted stars. By comparing our findings with data from the GALAH spectroscopic survey, we confirm that the observationally established diagnostics of [Al/Fe] vs. [Mg/Mn] also show a separation of in-situ and accreted stars in the simulation, but stars from different accretion events tend to overlap in this plane even without observational uncertainties. Looking at the relationship between stellar age and linear or logarithmic abundances, such as [Fe/H], we can clearly separate different groups of these stars if the uncertainties in their chemical makeup are less than 0.15 dex and less than 20\% for their ages. This method shows promise for studying the history of the Milky Way and other galaxies. Our work highlights how important it is to have accurate measurements of stellar ages and chemical content. It also shows how simulations can help us understand the complex process of galaxies merging and suggest how these events might relate to the differences we see between our Galaxy's thin and thick disk stars. This study provides a way to compare theoretical models with real observations, opening new paths for research in both our own Galaxy and beyond.
\end{abstract}



\section{Introduction}
\label{sec:intro}

The history of the Milky Way is a puzzle that has taunted astronomers for decades. The long lifetimes and rather conserved chemical composition of stars makes them key pieces in this puzzle, allowing us to use their star light as fossil record to gain insights into the historical processes that have led to our Galaxy as we know it today \citep{FreemanBlandHawthorn2002}. The advent of new large-scale stellar surveys, providing a wealth of information including measurements of elemental abundances\footnote{We define elemental abundances, such as [Fe/H], [Mg/Fe], and [Mg/Mn] as logarithmic ratios of two elemental number densities $N_\mathrm{X}$ and $N_\mathrm{Y}$ compared to the Sun, that is, $\left[\text{X/Y}\right]=\log_{10}\left(\frac{N_\mathrm{X}}{N_\mathrm{Y}}\right) -\log_{10}\left(\frac{N_\mathrm{X}}{N_\mathrm{Y}}\right)_\odot$.} for millions of stars, has enabled insights into the long and complex history of our Galaxy that have been previously unattainable \citep{Jofre2019}. Astrometric information for more than 1.5 billion stars from the \Gaia satellite \citep{Brown2016,Brown2018,Brown2021} has given us a new understanding of the structure of our Galaxy, while spectroscopic surveys like the Galactic Archaeology with HERMES (GALAH) Survey \citep{daSilva2015} or the Apache Point Observatory Galactic Evolution Experiment \citep[APOGEE,][]{Majewski2016} are complementing this vision with chemical fingerprints for millions of stars. Most notably, these surveys have led to the confirmation of a major accretion event around 8-10 billions years ago by \citet{Belokurov2018} and \citet{Helmi2018}. While this major merger of the \Gaia-Sausage-Enceladus (GSE) and the early Milky Way could explain the bimodality of stars in the Milky Way's disk, the causality is yet to be established beyond doubt. Our best avenue is to quantify the chemical and dynamical properties of accreted stars to answer outstanding questions like how much gas the GSE merger brought in to explain the significantly different chemistry of young versus old disk stars.

From our vantage point within the disk-dominated region of the Milky Way, it is difficult to identify the relatively low number of accreted stars. Intensive research has tested a variety of ways to find such stars - an extensive list of which is presented by \citet{Buder2022}. A first impulse was given by the series of papers by \citet{Nissen2010, Nissen2011, Nissen2014} and \citet{Schuster2012} who analysed the chemistry of stars in the kinematic halo of the Milky Way. They identified a population within the Galactic halo that showed significantly lower enhancements across a number of elements, hinting to an extra-Galactic origin. Starting from dynamic arguments, \citet{Belokurov2018} and \citet{Helmi2018} rediscovered this population as stars with outstanding eccentric and radial orbits and proved its significance via dynamical simulations and chemical evolution arguments. Especially the inference of merger masses via chemical patterns is only possible because of our ability to observe surviving dwarf galaxies \citep{Tolstoy2009, Helmi2018, Hasselquist2021, Carrillo2022}. Peaking further into the halo, \citet{Naidu2020} then showed that the halo is entirely comprised of substructure, including accreted stars. Studies of extragalactic sources can now also spatially resolve these source broadly and use age-metallicity relations to identify accreted stars \citep{Martig2021}, with tens of edge-on galaxies being observed by the MUSE integral field spectrograph through large programs like GECKOS \citep{GECKOS2023}. Such analyses are however limited by the fact that spatial and dynamical selections are less robust in a spatially and dynamically evolving galaxy over billions of years.

The key advantage of chemical tagging, when compared to dynamical tagging via velocities or energies, is that the composition of heavy elements, once locked in the atmosphere of a star, is expected to be much more conserved in a dynamically evolving galaxy. Building upon this premise, \citet{Hawkins2015}, \citet{Das2020} and \citet{Buder2022} assessed a number of potential abundance combinations for chemically tagging accreted stars. Among the many elements in the periodic table, they identified especially Na, Al, and Cu as exceptionally useful, if they can be measured. Using these elements together with the purer tracers of core-collapse supernovae (CCSN), like Mg, and supernovae type Ia (SNIa), like Mn, they found the abundance planes of [Na/Fe] vs. [Mg/Mn] and [Al/Fe] vs. [Mg/Mn] to reveal accreted stars in rather isolated loci.

Ages and abundances are, however, not directly observable and difficult to obtain from stellar spectra and any age-inference tool. Significant progress has been made in terms of the number and quality of observations, but we are still limited by our ability to observe a significant number of accreted stars with accurate and precise ages and abundances. While the GALAH survey can for example measure Na abundances, it is particularly difficult to measure Al abundances from its spectra. The observed stellar ages of GALAH DR3 are even more uncertain - with an average age uncertainty of $33\pm20\%$. Cosmological simulations, such as those used in this study, carry no age uncertainties and the particular simulation \citep{Buck2021} we use here, introduced in Sec.~\ref{sec:sim_data}, now also includes a wide range of chemical abundance information. 

Cosmological simulation projects such as \textsc{FIRE} \citep{Bonaca2017, Horta2023}, \textsc{EAGLE} \citep{Mackereth2019}, \textsc{NIHAO-UHD} \citep{Buck2020,Buck2021,Buck2023}, \textsc{AURIGA} \citep[][]{Grand2020}, \textsc{VINTERGATAN} \citep{Agertz2021} or  \textsc{HESTIA} \citep{Khoperskov2023} allow us to follow a self-consistent growth and chemical enrichment history, which is particularly important for metal-poor stars formed at redshifts $z>1$. Simulation groups have made further huge progress in reaching the numbers of stellar particles necessary to split datasets in phase-space and abundance space. The latest versions of cosmological simulations thus provide an exceptional data set for exploring the changing chemodynamic environment of our Milky Way - which is neither in equilibrium \citep[see e.g.][]{Antoja2018, BlandHawthorn2019} nor had an isolated evolution \citep{Helmi2020}. We can thus study the theoretically expected impact of massive mergers on the chemical evolution of Milky Way-like galaxies \citep{Buck2021,Buck2023} as well as our observational diagnostic tools without uncertainty margins to consider.

The aim of this project is to assess diagnostic tools for identifying accreted structures from observations in the Milky Way by comparing them with theoretical predictions from cosmological simulations. This assessment will be based on our ability to observe and quantify clear patterns and trends. In addition to testing chemical tagging, that is, relying only on abundances, we also test the use of age-abundance planes and make predictions on the necessary uncertainties for the diagnostic properties when assuming the Milky Way analogue as ground truth.

The paper is structured as follows: Sec.~\ref{sec:data} describes the data of GALAH observations and NIHAO simulations. Sec.~\ref{sec:comparison} performs an initial comparison of diagnostic plots, using different abundance planes. Sec.~\ref{sec:Age-abundance} considers how stellar abundances of accreted and in-situ stars change over time, and quantifies the results. Sec.~\ref{sec:discussion} discusses our findings and their implications. Sec.~\ref{sec:conc} concludes the paper.

\section{Data: Observations and simulations} \label{sec:data}

\subsection{Observational data from the GALAH Survey}\label{sec:obs_data}

Observations are provided through the GALactic Archaeology with HERMES (GALAH) survey. For our study, we use the abundance measurements from the third data release (DR3) of the GALAH Survey with its 588,571 stars \citep{Buder2021}. The main program of the survey targets nearby stars via their visual magnitudes (typically $9 < V < 12$ and $12 < V < 14$) while neglecting the Galactic plane ($\vert b \vert > 10\,\mathrm{deg}$). However, the data release also includes auxiliary programs of the K2 and TESS footprints \citep{Sharma2018, Sharma2019} as well as open and globular clusters \cite[for more details see][]{Buder2021}. Stars were targeted with the 2dF fibre optics positioner system at the Anglo-Australian Telescope \citep{Heijmans2012, Farrell2014} and the light of up to 400 stars was simultaneously fed into the High Efficiency and Resolution Multi-Element Spectrograph \citep[HERMES,][]{Barden2010, Sheinis2015}. HERMES delivers high-resolution ($R \sim 28,000$) spectra in four wavelength bands across the optical range which were reduced with the pipeline by \citep{Kos2017}.

Stellar parameters ($T_\text{eff}$, $\log g$, [Fe/H], $v_\text{mic}$, $v_\text{broad}$, and $v_\text{rad}$) and abundances for up to 31\footnote{Li*, C*, O*, Na*, Mg*, Al*, Si*, K*, Ca*, Sc, Ti, V, Cr, Mn*, Fe*, Co, Ni, Cu, Zn, Rb, Sr, Y, Zr, Mo, Ru, Ba*, La, Ce, Nd, Sm, and Eu, with * denoting elements calculated in 1D non-local thermodynamic equilibrium (non-LTE), instead of 1D LTE.} different elements across multiple nucleosynthesis channels are estimated using a modified version of the spectrum synthesis code Spectroscopy Made Easy \citep[\textsc{sme}][]{Valenti1996, Piskunov2017} and 1D \textsc{marcs} model atmospheres \citep{Gustafsson2008}. Eleven elements are computed in non-LTE \citep{Amarsi2020}, the others in local thermodynamic equilibrium (LTE), and additional constraints on the surface gravities are incorporated through inferred distances by \citet{BailerJones2021} and their limit on absolute magnitudes and luminosities.

In addition to the chemical information, we use the crossmatches of stars with value-added-catalog for stellar ages. The latter incorporates photometric and astrometric information from \Gaia eDR3 \citep{Lindegren2021a} and 2MASS \citep{Skrutskie2006} to estimate isochrone ages with the \textsc{bstep} code \citep{Sharma2018}.

We apply the following general quality and selection cuts to the data, firstly $\texttt{flag\_repeat} = 0$ to get one measurement per star, and secondly $\texttt{flag\_sp} = 0$, $\texttt{flag\_fe\_h} = 0$, $\texttt{flag\_Mg\_fe} = 0$, $\texttt{snr\_c2\_iraf} > 25$, and $\mathrm{[Fe/H]} > -2$ to get the most reliable stellar parameters while maintaining a sufficient amount of stars. Whenever we use additional elemental abundances of an element X, we also enforce $\texttt{flag\_X\_fe} = 0$.

While we have considered enforcing \textsc{bstep} age uncertainties below 50\%, we have decided against implementing a sharp limit, both to avoid complex influences\footnote{Of the 100\,307%
 stars above $8\,\mathrm{Gyr}$ 2\%%
, 10\%%
, and 37\%%
 have uncertainties above $50\%$, $40\%$, and $30\%$, respectively. Of the 184\,616%
 stars below $8\,\mathrm{Gyr}$ already 19\%%
 have uncertainties above $50\%$.} in the selection function and because it does not significantly impact the stars with ages above $8\,\mathrm{Gyr}$.

To allow better comparability with the simulations later on (see Sec.~\ref{sec:location}), we also limit our sample to stars with Galactic latitude $\vert b \vert > 10\,\mathrm{deg}$ and distance $D_\varpi < 4.2\,\mathrm{kpc}$ \citep[based on photogeometric distances from][]{BailerJones2021}, the 95th distance percentile of GALAH DR3 \citep{Buder2021}. After applying all these cuts, our sample still consists of 284\,923
stars, allowing a statistically meaningful comparison with theoretical predictions.

\subsection{Theoretical predictions from a NIHAO Zoom-in simulation}\label{sec:sim_data}

The model perspective on accretion is provided through a cosmological zoom-in simulation of a Milky Way analogue (\texttt{g8.26e11}) from the \textit{Numerical Investigation of a Hundred Astronomical Objects} \citep[NIHAO,][]{Wang2015}. The model galaxy has a total mass (gas, stars and dark matter) of $8.26 \cdot 10^{11}\,\mathrm{M_\odot}$ and contains $4 \cdot 10^{10}\,\mathrm{M_\odot}$ stellar mass with a stellar mass resolution of $1.06 \cdot 10^{5}\,\mathrm{M_\odot}$ \citep{Buck2021} and was calculated as part of the NIHAO-UHD project \citep{Buck2020b}.

Simulations were carried out with the smoothed particle hydrodynamics code \texttt{Gasoline2} \citep{Wadsley2017} using cosmological parameters from \citet{Planck2014} with initial conditions and energetic feedback descriptions from the NIHAO project \citep{Wang2015}. Zoom-in simulations were then performed as described in detail by \citet{Buck2021} with star formation following \citet{Stinson2006} and energetic feedback following \citet{Stinson2013}.

Because computational resources still limit the mass resolution of simulations, we are relying on tracer particles that represent simple stellar populations (SSPs) with the same age, overall metallicity and discrete initial mass function (IMF). \citet{Buck2021} have implemented the flexible chemical evolution code \textsc{chempy} \citep{Rybizki2017} to calculate the chemical yields for the SSPs. In particular, we use the alternative (\texttt{alt}) setup of \textsc{chempy} that assumes a \citet{Chabrier2003} IMF with high-mass slope of $\alpha_\text{IMF} = -2.3$ over a mass range of $0.1-100\,\mathrm{M_\odot}$ for SSPs across a metallicity range of $Z/Z_\odot \in [10^{-5},2]$. The code calculates the contribution from asymptotic giant branch (AGB) stars, CCSN across a mass range of $8-40\,\mathrm{M_\odot}$, and SNIa with a an exponential function with exponent $-1.12$, a delay time of $40\,\mathrm{Myr}$, and a normalization of the SNI rate of -2.9. For each of these nucleosynthetic channels, yields from the following studies are used: \citet{Limongi2018} for CCSN, \citet{Seitenzahl2013} for SNIa, and \citet{Karakas2016} for AGB stars.

To achieve a roughly similar selection as the observational data (see Secs.~\ref{sec:obs_data} and \ref{sec:location}), while maintaining enough star particles, we limit the simulated data set to star particles within a torus at $8.2\,\mathrm{kpc}$ galactocentric radius with a $4.2\,\mathrm{kpc}$ tube radius (corresponding to the 95th distance percentile of the GALAH sample) and neglect star particles within $\vert b \vert~<~10\,\mathrm{deg}$ or $\mathrm{[Fe/H]} < -2$. We use the last simulation snapshot at a cosmic time of $13.8\,\mathrm{Gyr}$ to estimate the ages from the reported formation time.

The simulation traces the elemental abundances of H, He, C, N, O, Ne, Mg, Al, Si, P, S, V, Cr, Mn, Fe, Co, and Ba. Of these, ten correspond with the GALAH data and allow us to compare diagnostic plots possible with abundances of C, O, Mg, Al, Si, V, Cr, Mn, Co and Ba. To allow a better comparison of simulated to observed abundance patterns, we shift the abundance patterns [X/H] by the values listed in Table~\ref{tab:shift_4to5gyr}, such that the median abundance for [X/H] is Solar for the stars with ages of Solar age ($4-5\,\mathrm{Gyr}$) in the GALAH-like footprint of the Milky Way analogue.

The simulation records galactocentric Cartesian positions $(X,Y,Z)$ and velocities $(V_X,V_Y,V_Z)$ for each particle, which we transform into a galactocentric Cylindrical coordinate frame via
\begin{align}
    R = \sqrt{X^2 + Y^2} \qquad &\& \qquad V_R = \frac{X*V_X + Y*V_Y}{R} \\
    \phi = \arctan(X/Y) \qquad &\& \qquad V_\phi = \frac{-Y*V_X + X*V_Y}{R} \\
    z = Z  \qquad &\& \qquad V_z = V_Z,
\end{align}
while using the \textsc{numpy} \textsc{arctan2} package.

\begin{table}
    \centering
    \caption{
    \textbf{Median abundance [X/H] for particles of the Milky Way analogue with $4 < \mathrm{Age~/~Gyr} < 5$ in the observation footprint.}
    These were used to shift the the abundance to agree with an expected Solar value.
    }
    \begin{tabular}{cccccc}
\hline \hline
Element & Shift & Element & Shift & Element & Shift \\
\hline
{[C/H]} & -0.27 & {[N/H]} & -0.39 & {[O/H]} & -0.26 \\
{[Ne/H]} & -0.20 & {[Mg/H]} & 0.17 & {[Al/H]} & 0.08 \\
{[Si/H]} & -0.34 & {[P/H]} & -0.21 & {[S/H]} & -0.29 \\
{[V/H]} & -0.02 & {[Cr/H]} & -0.19 & {[Mn/H]} & -0.36 \\
{[Fe/H]} & -0.15 & {[Co/H]} & -0.03 & {[Ba/H]} & -0.70 \\
\hline \hline
\end{tabular}

    \label{tab:shift_4to5gyr}
\end{table}


\section{Diagnostic plots: Observation vs. Simulation}
\label{sec:comparison}

In Sec.~\ref{sec:intro}, we have reviewed several observation-based plots that have been used to identify accreted stars in the Milky Way. In this section, we are now assessing if the simulated Milky Way analogue shows similar signatures when using these diagnostic plots. In Sec.~\ref{sec:location}, we first analyse the major diagnostic plots ([Fe/H] vs. [Si/Fe], [Al/Fe] vs. [Mg/Mn], and Age vs. [Fe/H]) while focusing on their difference for the full galaxy and a GALAH-like footprint of it. We then focus on the general abundance trends with iron abundance in Sec.~\ref{sec:feh_xfe}, before zeroing in on the abundance-abundance plane of [Al/Fe] vs. [Mg/Mn] in Sec.~\ref{sec:alfe_mgmn}.

\begin{figure}
	\includegraphics[width=\columnwidth]{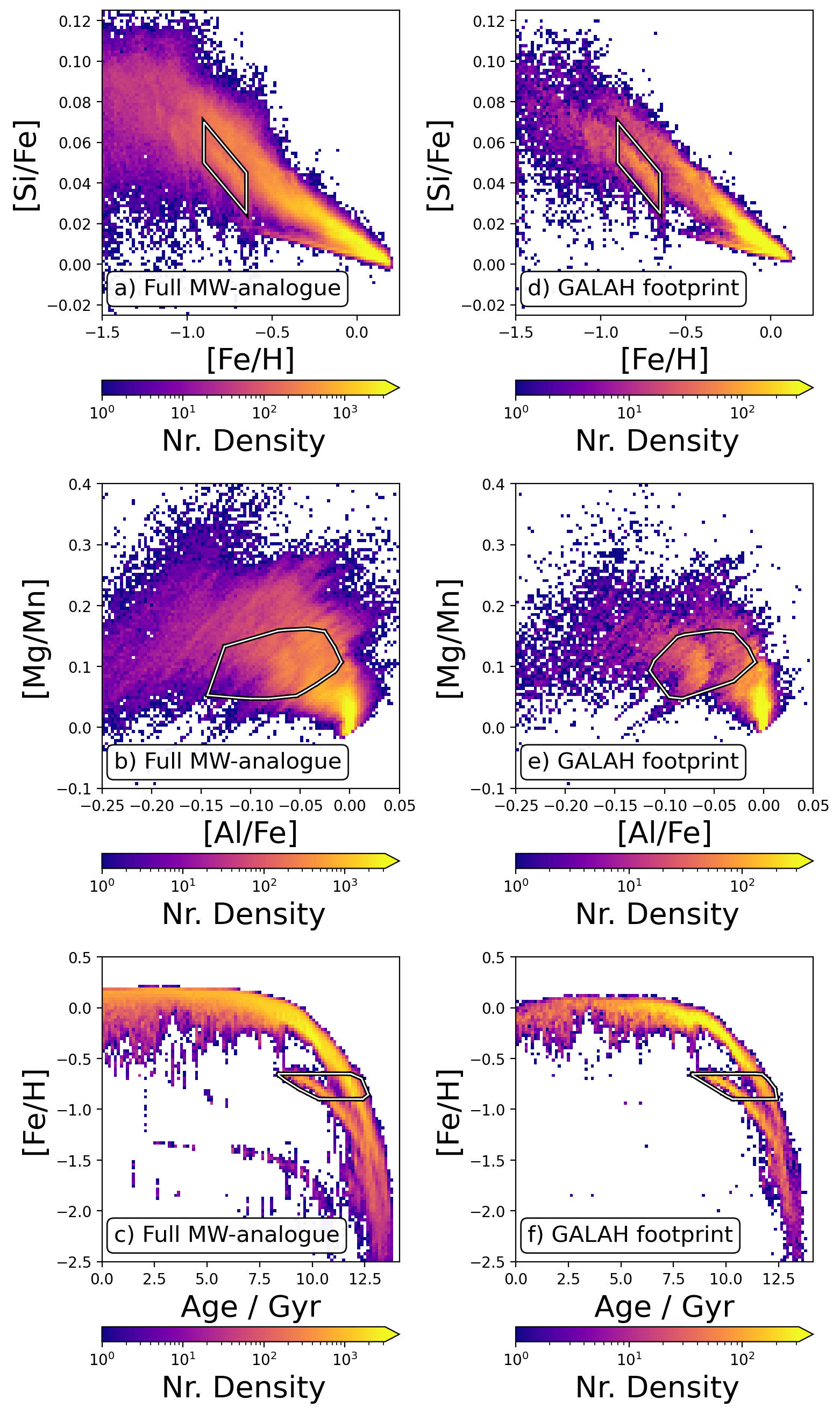}
    \caption{
    \textbf{Density distribution of all particles (left panels) and those in a GALAH-like footprint (right panels) for particles of the NIHAO simulation \texttt{g8.26e11} (color-coded by logarithmic density).}
    \textbf{Panels a) and d):} iron vs. silicon abundances as tracers of contribution from SNIa and CCSN. \textbf{Panels b) and e):} abundance combination [Al/Fe] vs. [Mg/Mn] used to trace accreted stars in observations.
    \textbf{Panels c) and f):} age-[Fe/H] relations.  
    The selection in panels a) and e) are drawn with a polygon with vertices at (-0.9,0.05),(-0.65,0.025),(-0.65,0.045), and (-0.9,0.07). A convex hull polygon of the selected stars is then drawn in other panels.}
    \label{fig:low_alpha_halo}
\end{figure}

\subsection{Location matters: Full galaxy versus Solar neighborhood} \label{sec:location}

Naïvely, we initially plot the diagnostic plots of observers, namely the ratio of CCSN to SNIa via [Si/Fe] as function of iron abundance [Fe/H] in Fig.~\ref{fig:low_alpha_halo}a as well as the previously successful selection plane of [Al/Fe] vs. [Mg/Mn] in Fig.~\ref{fig:low_alpha_halo}b for the complete Milky Way analogue. In the absence of observational noise, it could be expected that areas of in-situ and accreted stars might clearly separate. To the contrary, the [Al/Fe] vs. [Mg/Mn] for the whole galaxy show seemingly even more scatter than the observations when inspecting the whole galaxy. Tracing the selection of accreted stars, that is, the stars with lower [Si/Fe] around $-0.90 < \mathrm{[Fe/H]} < -0.65$ with a selection box, does not lead to a unique location in the [Al/Fe] vs. [Mg/Mn] in Fig.~\ref{fig:low_alpha_halo}b when showing the extend of these stars with a convex hull outline. While this may seem surprising for observers at first sight, it has been seen previously also in the \textsc{Auriga} \citep[][]{Grand2020, Orkney2022}, \textsc{HESTIA} \citep[][see their Fig.~12]{Khoperskov2023c},  as well as the \textsc{VINTERGATAN} \citep[][see their Fig.~A2]{Rey2023} simulations, when tracing different merger subpopulations.

When applying a similar selection to NIHAO as would be expected from GALAH observations in the Solar vicinity (see Sec.~\ref{sec:sim_data}), both the [Fe/H] vs. [Si/Fe] plane in Fig.~\ref{fig:low_alpha_halo}d as well as the [Al/Fe] vs. [Mg/Mn] plane in Fig.~\ref{fig:low_alpha_halo}e cleared up significantly, with the accreted sequence becoming more prominent around $\mathrm{[Al/Fe]}$ of -0.3.

\begin{figure*}
	\includegraphics[width=0.94\textwidth]{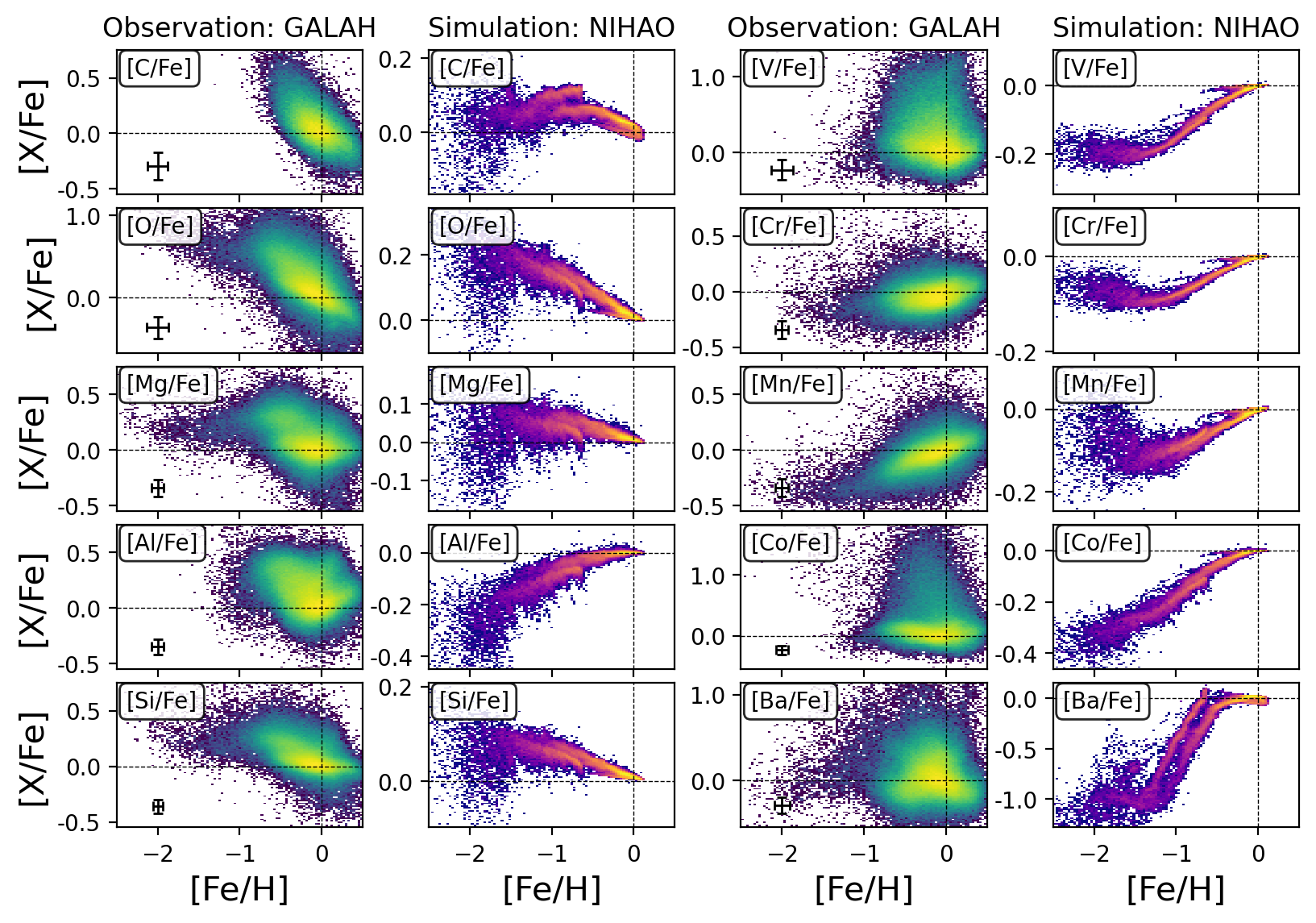}
    \caption{
    \textbf{Logarithmic density distribution of elemental abundances [X/Fe] versus [Fe/H] for the ten elements overlapping between GALAH and NIHAO.} 
    \textbf{First and third rows} show observed data from GALAH while \textbf{second and fourth rows} show the simulated data for the same abundance ratios in a GALAH-like footprint. Brighter colors (towards yellow) indicate higher densities. For better visibility, plot ranges are adjusted to the data rather than to be equal among each panel. Dashed lines indicate Solar values. Simulated abundance trends for N, Ne, P, and S are shown in Fig.~\ref{fig:appendix_feh_xfe_nneps}.}
    \label{fig:FeH_XFe}
\end{figure*}

Intrigued by the difference in clarity of the chemical planes, we also show the two age-metallicity relations - or more accurately age-[Fe/H] planes - for the full Milky Way analogue in Fig.~\ref{fig:low_alpha_halo}c and the observational footprint in Fig.~\ref{fig:low_alpha_halo}f. The clearly arising separations of sequences in both selections motivates a more dedicated analysis of this plane, as will follow in Sec.~\ref{sec:Age-abundance}. The tight sequences, however, also raise the question, if our Milky Way analogue is indeed similar to our Milky Way, which will be discussed in Sec.~\ref{sec:discussion_how_comparable_are_nihao_and_milkyway}. Under this section's theme of identifying differences in the planes of full and limited sample selection, we notice a slightly larger scatter of abundances in Fig.~\ref{fig:low_alpha_halo}c, that is, a differences of iron abundances outside the Solar vicinity, as well as the disappearance of a metal-poor ($\mathrm{[Fe/H]} \sim -1.5$) sequence below $10\,\mathrm{Gyr}$ when limiting the selection to the GALAH footprint.

\textbf{Key Takeaway:} Location matters, or better: a clever sample selection - even for simulations!

\subsection{Abundance trends: [Fe/H] vs. [X/Fe]} \label{sec:feh_xfe}

In Fig.~\ref{fig:FeH_XFe}, we show the elemental abundances [X/Fe] as a function of iron abundance [Fe/H] of the 10 elements X that overlap between GALAH observations and NIHAO simulations. To first order we see an agreement of general trends for most elements, that is, elements that are produced by CCSN like O, Mg, and Si are showing a high plateau for low [Fe/H] and a decrease towards higher iron abundances - as expected by the onset of Fe-producing SNIa. The iron-peak elements Cr as well as Mn and Co show increasing abundances from low values at low [Fe/H] towards solar values at solar [Fe/H]. Abundance estimates for V and Co do not agree, which is partially (but not fully) caused by unreliably high measurements of [V/Fe] and [Co/Fe] in GALAH \citep{Buder2021}. For Ba, we note tightly increasing enhancements towards Solar [Fe/H], starting from initially low values around $\mathrm{[Ba/Fe]} \sim -0.5$. Furthermore, we notice a clear second (accreted) sequences below $\mathrm{[Fe/H]} < -0.5$ for C as well as Ba. While not shown in this manuscript, we note that when plotting the full simulated galaxy in Fig.~\ref{fig:FeH_XFe}, no significant difference in trends can be identified, but a larger spread of abundances - most notably towards the lowest metallicities.

When focusing on the actual absolute abundances (note the difference in y-axis), we note significant differences between the chemical evolution sequences. To understand this difference, we first need to remind ourselves that the observational data is limited by the ability to measure abundances (note the missing abundances of [C/Fe] for low metallicities due to detection limits) and includes measurement noise, while the theoretical data does not necessarily have to reproduce the formation scenario of the Milky Way and chemical enrichment tracing is limited to our best current understanding of nucleosynthesis, see e.g. the discussion of nucleosynthetic yields in \citet{Buck2021}.

That said, without adjusting for abundance-offsets, we note that none of the abundance sequences in the Solar metallicity regime of $\mathrm{[Fe/H]} \sim 0$ intercept with the expected Solar value. While C, O, Si, Cr, Mn, and Ba are significantly overproduced between $0.05$ (for Cr) and $0.5\,\mathrm{dex}$ (for Ba), we notice a significant under-abundance on the $-0.3..-0.1\,\mathrm{dex}$ level for Mg, Al, V, and Co. Nevertheless, for the remainder of this manuscript this poses no severe limitation as we are mainly interested in a differential signal in order to separate in-situ and accreted stars.

\textbf{Key Takeaway:} While simple abundance-metallicity plots do not easily reveal accreted structures for most elements, both C and Ba abundances show the potential to identify different abundance sequences - if the nucleosynthetic processes of the simulation are accurate.

\subsection{Abundance-abundance plots: [Al/Fe] vs. [Mg/Mn]} \label{sec:alfe_mgmn}

We now turn to abundance-abundance plots of [Al/Fe] vs. [Mg/Mn] that were already mentioned in Sec.~\ref{sec:location}. These plots combine the enrichment channels of CCSN vs. SNIa, that is [Mg/Mn], and the enrichment of Al with more complex metallicity-dependent enrichment pathways \citep{Hawkins2015, Das2020, Kobayashi2020}. The enhancements of these elements should differ significantly when comparing in-situ stars born in the early Milky Way with the less massive system that it accreted because of the expected lower star formation intensity (and thus iron abundance) and occurrence of CCSN in the latter. 

\begin{figure}
	\includegraphics[width=\columnwidth]{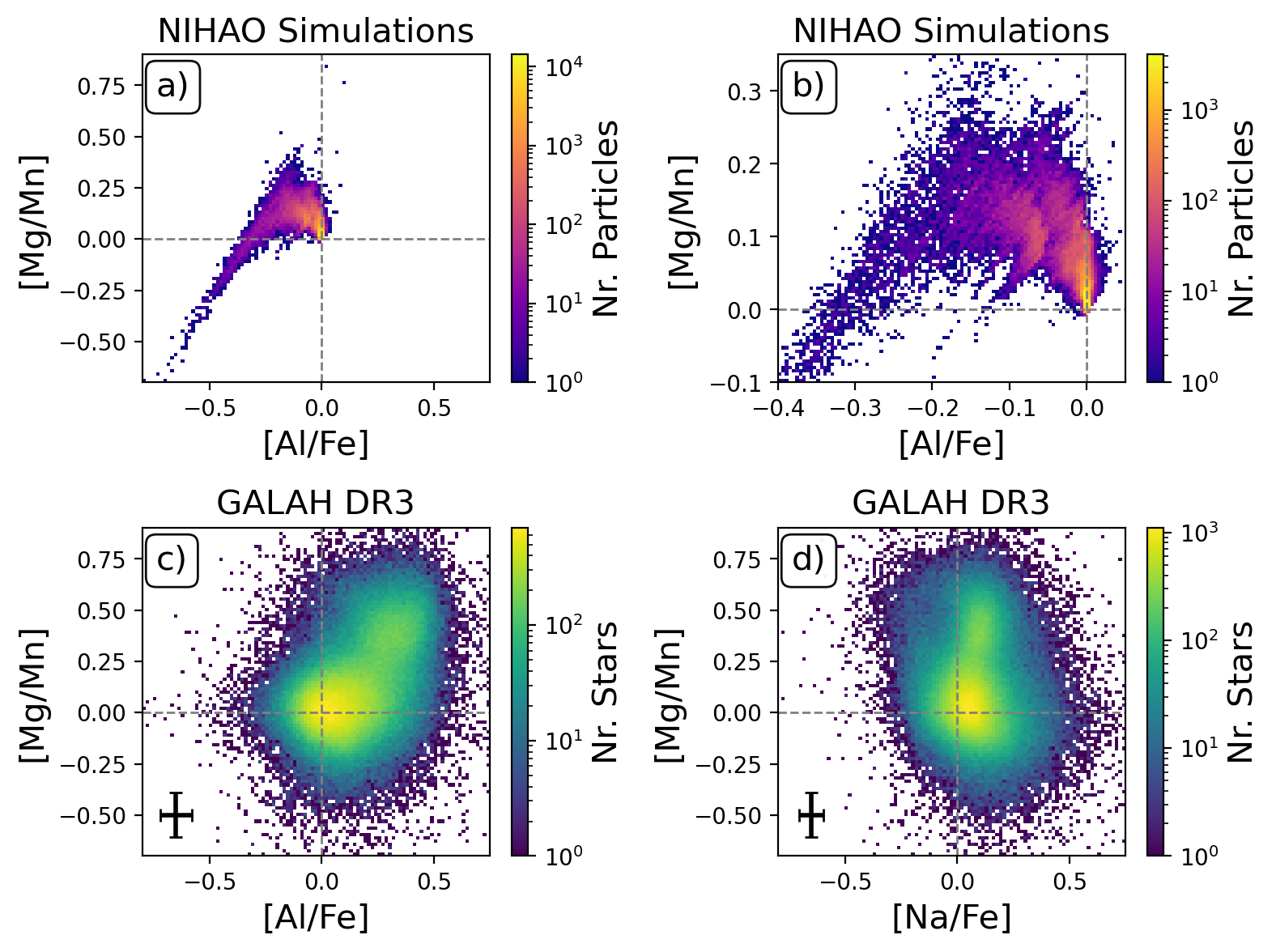}
    \caption{
    \textbf{Comparison of the diagnostic plot of odd-Z elements [Al/Fe] or [Na/Fe] vs. the ratio of CCSN and SNIa elements [Mg/Mn] in simulation (top) and observation (bottom), with accreted stars being expected in the upper left of the distributions.}
    \textbf{Left panels a and c} use the same axis limits, whereas \textbf{panel b} is a zoomed version of the simulation. \textbf{Panel d} complements the picture, as [Al/Fe] is hard to measure in GALAH (see missing data points in top left of panel c).
    }
    \label{fig:NaFe_MgMn_selection_Age_FeH_dissection}
\end{figure}

When using the same axis limits for simulation and observation (Figs.~\ref{fig:NaFe_MgMn_selection_Age_FeH_dissection}a and c), we notice the strong offset and significantly smaller spread of abundances in the simulation, in agreement with our previous findings for abundances in Sec.~\ref{sec:feh_xfe}. When zooming into the relevant [Al/Fe] vs. [Mg/Mn] regime for the simulations, however, we notice a better qualitative agreement, in that we see two significant overdensities for simulated and observed galaxy. For the observations, we identify the lower and upper overdensity at [Mg/Mn] of 0.0 and 0.4. These have previously been confirmed as low- and high-$\upalpha$-sequence (or thin and thick disk), respectively \citep{Hawkins2015, Buder2022}. These sequences show similar Solar or even enhanced [Na/Fe] and [Al/Fe]. In contrast to this, the major overdensity in the simulation is found around (-0.25,-0.5) with tails reaching higher [Mg/Mn]. The minor overdensity is found towards higher [Mg/Mn] as well as lower [Al/Fe] - coinciding with the expected region of accreted stars \citep{Horta2021} which are less numerous in the observational data of GALAH DR3.

\begin{figure*}
	\includegraphics[width=0.995\textwidth]{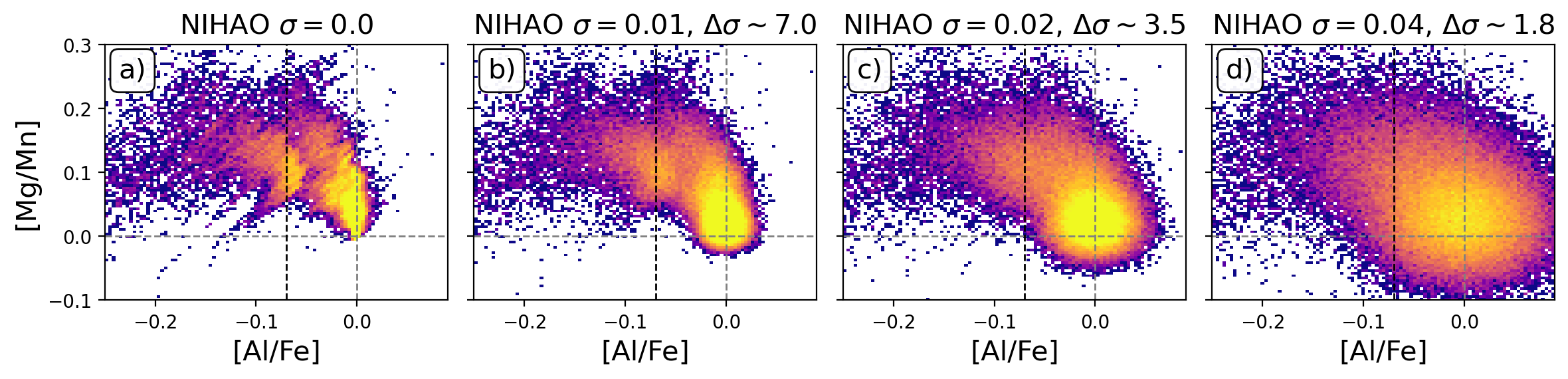}
    \caption{
    \textbf{Logarithmic density distribution of particles in the [Al/Fe] vs. [Mg/Mn] plane when adding increasing levels of noise to the observational footprint of the NIHAO simulation.} Gray dashed lines indicate the Solar value, whereas the black dashed line indicates the visual centre in  [Al/Fe] of the accreted sequence at $-0.07\,\mathrm{dex}$. Panel titles indicate the added noise level, relating to a decrease of separation significance in [Al/Fe] down to $\Delta \sigma \sim 1.8$ for panel d). Color maps have been saturated at maximum values of 100 for all panels for increased contrast and better comparability.
    }
    \label{fig:mgmn_alfe_with_noise}
\end{figure*}

To achieve a more realistic comparison, we add noise on the order of $0.01$ and $0.04\,\mathrm{dex}$ to the [Al/Fe] vs. [Mg/Mn] plane of the NIHAO simulation in Fig.~\ref{fig:mgmn_alfe_with_noise}. It has to be noted though, that this noise has to be compared to the separation of accreted and in-situ sequence, which is on the order of $0.07\,\mathrm{dex}$ for [Al/Fe] in the simulation (see black dashed lines in Fig.~\ref{fig:mgmn_alfe_with_noise}), compared to $0.18\,\mathrm{dex}$ in the Milky Way \citep[][see their Tab.~4]{Buder2022}. This means that a noise of $0.03\,\mathrm{dex}$ ($\Delta \sigma \sim 2.3$) in NIHAO has a similar effect as the noise of $0.08\,\mathrm{dex}$ in GALAH DR3. When comparing the different panels, we see that the initially very clear separation of the accreted and in-situ sequence quickly washes out to the point of current observational noise (between Figs.~\ref{fig:mgmn_alfe_with_noise}c and \ref{fig:mgmn_alfe_with_noise}d), where more sophisticated methods are needed to assign particles to either sequence, as done in the past by observers \citep{Das2020, Buder2022}.

\textbf{Key Takeaway:} The abundance-abundance plots of [Al/Fe] vs. [Mg/Mn] show qualitative agreement with the observational data, but with significant abundance offsets and a more pronounced overdensity of accreted stars than the observations in the absence of noise. This could indicate a more massive merger progenitor than the Milky Way's GSE or hint at our inability to actually measure Mg, Mn, and Al or Na abundances of accreted stars as well as reflect theoretical uncertainties in the yields of Mg, Mn, Al and Na. When adding realistic uncertainties to this plane, the separation washes out to the point where an assignment to either sequence needs more advanced methods.

\section{A new angle: age-abundance-distributions}\label{sec:Age-abundance}

Fortunately, chemistry is not the only conserved property of stars. We therefore turn to stellar ages and assess age-abundance trends in the data, starting from the age-[Fe/H] relationship that we already showed in Figs.~\ref{fig:low_alpha_halo}c and f. This relation has proven to be a powerful tool not only for resolving processes with observed data of our Milky Way \citep[e.g.][]{Twarog1980, Edvardsson1993, Nordstroem2004, Casagrande2011, Feuillet2019, Xiang2022}, but also other galaxies - for example to identify accreted stars \citep[e.g.][]{Pinna2019b, Martig2021}. However, observations still suffer from larger uncertainties on the order of $15-50\%$ for most stars, so examining simulation plots removes this barrier and allow us to determine chemical enrichment as a function of age. Additionally, turning towards stellar ages in the simulations focuses on the fundamental variable age and removes uncertainties associated with uncertain yield sets.

\subsection{Clearly separated sequences in the age-[Fe/H] relation of the Milky Way analogue} \label{sec:age-feh-sequences}

Based on Fig.~\ref{fig:low_alpha_halo}f, we identify two definite and one tentative sequence in the age-metallicity relation with the following selections, selecting 75\,440 particles for Sequence 1, 13\,398 particles for Sequence 2, and 133 particles for Sequence 3 of the 88\,980 particles in the footprint%
:
\begin{align}
    \text{Sequence~1} &= \begin{cases}
        \mathrm{[Fe/H] > -0.2} \text{ or} \\
        (\mathrm{[Fe/H]} > -1.0 \text{ and } \mathrm{age} < 7\,\mathrm{Gyr}) \text{ or} \\
        10^{\mathrm{[Fe/H]} + 0.35} > 2.0 - 0.15 \cdot \mathrm{age}~/~\mathrm{Gyr}
    \end{cases} \label{eq:sequence1} \\
    \text{Sequence~2} &= \begin{cases}
        \text{Not Sequence~1 and }\mathrm{Age} > 6\,\mathrm{Gyr} \text{ and} \\
        10^{\mathrm{[Fe/H]} + 0.45} > 1.87 - 0.15 \cdot \mathrm{age}~/~\mathrm{Gyr}
    \end{cases}  \label{eq:sequence2} \\
    \text{Sequence~3} &= \text{Not Seq.~2 and not Seq.~3 and } \mathrm{age} > 7~/~\mathrm{Gyr}  \label{eq:sequence3}
\end{align}

\begin{figure*}
	\includegraphics[width=\textwidth]{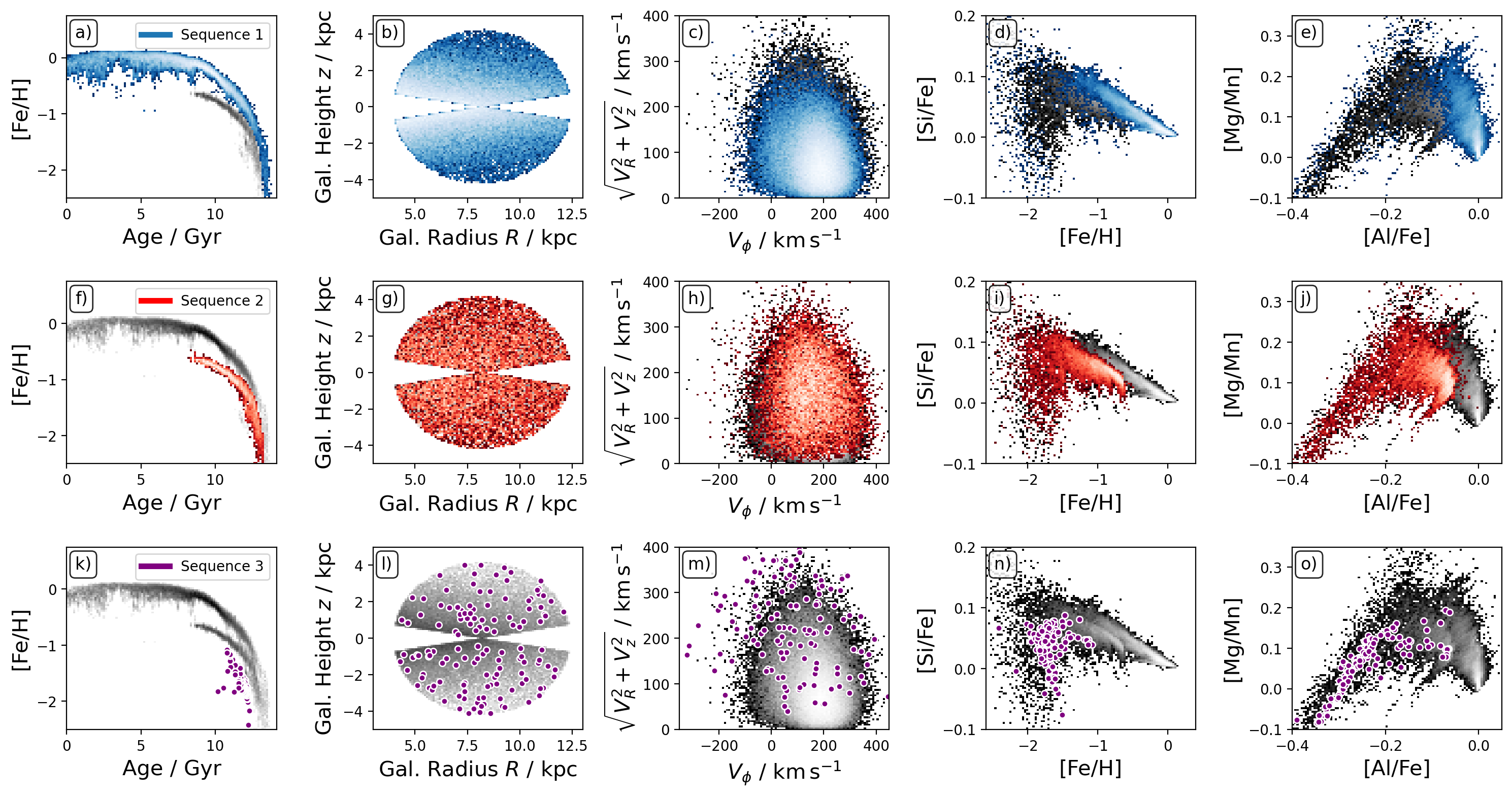}
    \caption{
    \textbf{Logarithmic density distributions of chronochemodynamic properties of the three separate sequences identified in the age-[Fe/H] relation (first column) - blue for Sequence 1 in the top row, red for Sequence 2 in the middle row, and purple for Sequence 3 in the bottom row.
    }
    \textbf{The second column} shows the spatial density of stars in $R$-$z$.
    \textbf{The third column} shows the kinematic extend of stars in the Toomre diagram of $V$ vs. $\sqrt{U^2+W^2}$.
    \textbf{The fourth and fifth columns} show how the chemical distribution in the [Fe/H] vs. [Si/Fe] and [Al/Fe] vs. [Mg/Mn] plane, respectively.
    }
    \label{fig:three_sequences_traced}
\end{figure*}

We discuss these numbers in comparison with the Milky Way later in Sec.~\ref{sec:discussion_how_comparable_are_nihao_and_milkyway}. In this section, we are primarily interested in tracing the three sequences across their chronochemodynamic properties, as done with separate rows in Fig.~\ref{fig:three_sequences_traced}, starting with the age-[Fe/H] relation.

For Sequence~1, the most metal-rich sequence with 85\%%
 of particles in the footprint, we see the stars being distributed closest to the galactic plane ($\vert Z \vert < 2\,\mathrm{kpc}$) in Fig.~\ref{fig:three_sequences_traced}b. Their kinematics in Fig.~\ref{fig:three_sequences_traced}c) likens a Toomre diagram of the Milky Way  \citep[][see their Fig.~1a]{Helmi2018} but with a smoother transition of thin and thick disk. The smooth transition of thin and thick disk in kinematic space could lead to the question if the Milky Way analogue truly has a thick disk \citep[see][who raised the same question for the actual Milky Way]{Bovy2012b}. This impression is confirmed in chemical space of [Fe/H] vs. [Si/Fe] in Fig.~\ref{fig:three_sequences_traced}d with its smooth sequence from higher $\mathrm{[Si/Fe]} \sim 0.3$ at $\mathrm{[Fe/H] \sim -1}$ to $\mathrm{[Si/Fe]} \sim 0.2$ at $\mathrm{[Fe/H] \sim 0}$. Similarly, we see a smooth distribution with highest $\mathrm{[Al/Fe]} \sim -0.22$ in the [Al/Fe] vs. [Mg/Mn] plane. We refrain from further dissecting this population in old and young sequence to establish a thin and thick disk as this is not the focus of this work.

Sequence~2 with 15\%%
 of particles in the footprint shows the behaviour of kinematic halo stars with a spatial distribution not restricted to the galactic plane (Fig.~\ref{fig:three_sequences_traced}g). While \citet{Buder2022} found $V_\phi = {20}_{-70}^{+100}\,\mathrm{kpc\,km\,s^{-1}}$ when selecting the major accretion remnant in the Milky Way via chemical compositions, we find significantly prograde orbits with $V_\phi = 150_{-100}^{+100}\,\mathrm{kpc\,km\,s^{-1}}$%
 (Fig.~\ref{fig:three_sequences_traced}h), suggesting a net rotation almost as large as the rotation of sequence 1 ($V_\phi = 180_{-80}^{+60}\,\mathrm{kpc\,km\,s^{-1}}$%
)\footnote{Note though, the net rotation of this population is strongly connected to the specific orbit of the merging dwarf galaxies and we do not expect any close agreement with specific Milky Way values here.}. Once again, we notice the offset towards lower abundance enhancement of this sequence in the [Fe/H] vs. [Si/Fe] plane as well as the [Al/Fe] vs. [Mg/Mn] plane with respect to sequence 1 (Figs.~\ref{fig:three_sequences_traced}i and j, respectively).

Finally, we notice that the smallest Sequence~3 (only 0.1\%%
 of the stars in the footprint) significantly overlaps with Sequence~2 in all of the chemodynamic properties. In addition to the very clear separation in age-abundance space (Fig.~\ref{fig:three_sequences_traced}k), the only way to identify particles of this sequence would be through their relative overdensities in the abundance-abundance planes (Figs.~\ref{fig:three_sequences_traced}n and o).

\textbf{Key Takeaway:} We find a smooth transition of thin and thick disk (identified as Sequence 1) of the Milky Way analogue in Figs.~\ref{fig:three_sequences_traced}a-e. Sequence 2 clearly presents itself as an accreted sequence in any of the plots of Figs.~\ref{fig:three_sequences_traced}f-j. In particular, the dominance of the thin/thick disk makes it hard to identify halo stars in spatial and kinematic space, wheres age-abundance and abundance-abundance plots show less contamination. Finally, we note that the smallest sequence (with lowest [Fe/H] values) is overlapping with the major accreted sequence spatially, kinematically, and chemically, and can only be separated as clump in the age-abundance (Fig.~\ref{fig:three_sequences_traced}k) or abundance-abundance space (Figs.~\ref{fig:three_sequences_traced}n and o). Having established the age-abundance planes as the best diagnostic plane (in the absence of measurement uncertainties), we are now concerned with the information held in age-abundance relations beyond the age-[Fe/H] relation.

\subsection{Age-abundance relations beyond age-[Fe/H]} \label{sec:age-abundance-sequences}

Given the various abundance distributions in the [Fe/H] vs. [X/Fe] space from Fig.~\ref{fig:FeH_XFe}, we are now interested in the trends for different elements with age. Previous research has suggested that not necessarily [X/Fe], but [X/H] may be a better tracer of accreted stars \citep[e.g.][for {[Mg/H]}]{Fuhrmann2017, Feuillet2021}. Furthermore, we are inspired by the work by \citet{Weinberg2019, Weinberg2021} as well as \cite{Griffith2019, Griffith2022}, who modeled nucleosynthetic contributions by SNIa and CCSN with amplitudes such as $A_\text{cc} = 10^\mathrm{[Mg/H]}$ \citep{Weinberg2019}. To avoid confusion with both the abundance notation $A(\mathrm{X})$ and their notation $A_\mathrm{cc}$/$A_\mathrm{Ia}$, we want to explicitly lay out our notation of rewriting the element abundances in a linear way as $N_\mathrm{X/H}$, so that they are linearly proportional to the ratio of number densities $N_\mathrm{X}/N_\mathrm{H}$:
\begin{align}
    \left[\text{X/H}\right] \varpropto \log_{10}\left(\frac{N_\mathrm{X}}{N_\mathrm{H}}\right) \quad \Rightarrow \quad N_\mathrm{X/H} = 10^{\mathrm{[X/H]}} \varpropto \frac{N_\mathrm{X}}{N_\mathrm{H}}
\end{align}

\begin{figure*}
	\includegraphics[width=\textwidth]{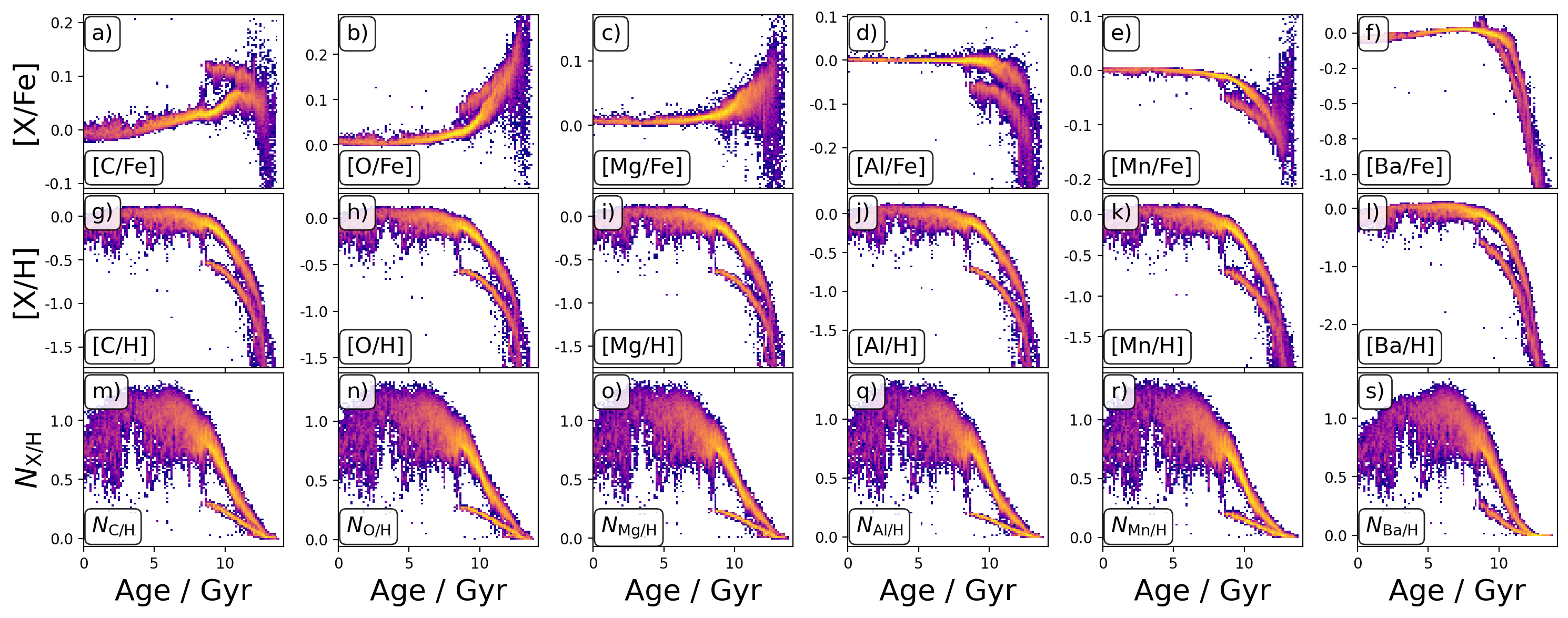}
    \caption{
    \textbf{Age-Abundance trends of the simulated Milky Way analogue in the observation footprint}.
    \textbf{Columns} show the different the elements C, O, Mg, Al, Mn, and Ba (from left column to right column), respectively.
    \textbf{Rows} show different notations of abundances, that is, logarithmic abundance {[X/Fe]} relative to iron (top row), logarithmic abundance {[X/H]} relative to hydrogen (middle row), and linear number density ratio {$N_\mathrm{X/H}$} (bottom row).
    }
    \label{fig:age_xfe_xh_nxh}
\end{figure*}

\subsubsection{Age-{$\mathrm{[X/Fe]}$} or age-{$\mathrm{[X/H]}$} or age-{$N_\mathrm{X/H}$?}} \label{sec:age_xfe_xh_nxh}

In Fig.~\ref{fig:age_xfe_xh_nxh} we show the age-abundance distributions for six of the elements overlapping between NIHAO and GALAH, namely C, O, Mg, Al, Mn, and Ba. The other abundance trends are appended in Fig.~\ref{fig:appendix_xfe_xh_nx} as their trends liken those already shown in Fig.~\ref{fig:age_xfe_xh_nxh} and they thus do not add new insights.

First, we discuss the top panels of Fig.~\ref{fig:age_xfe_xh_nxh} showing the age-[X/Fe] trends. As pointed out in our description of Fig.~\ref{fig:FeH_XFe}, the NIHAO simulation show a rather flat, enhanced [C/Fe] abundance for accreted stars, that is, $\mathrm{[C/Fe]} \sim 0.11$ compared to  $\mathrm{[C/Fe]} < 0.07$ at around $10\,\mathrm{Gyr}$. For O in Fig.~\ref{fig:age_xfe_xh_nxh}b, we notice a decrease of high [O/Fe] abundance with decreasing stellar age and an emerging bifurcation with higher $\mathrm{[O/Fe]} \sim 0.1$ for accreted stars. For Mg, the in-situ and accreted components overlap in the age-[Mg/Fe] plane (Fig.~\ref{fig:age_xfe_xh_nxh}c). For both [Al/Fe] and [Mn/Fe], we see $~0.05\,\mathrm{dex}$ lower abundance for accreted stars than for in-situ stars, with a clearer separation of sequences for [Al/Fe] at oldest ages. Finally, the [Ba/Fe] plane shows a lower initial abundance of [Ba/Fe] at oldest ages, which is expected - at least qualitatively - due to the delay of Ba-production by AGB stars \citep{Karakas2016}. We note here that the production of Ba relative to Fe for accreted stars actually surpasses those of in-situ ones, as can be seen from the crossing of Sequence 1 and 2 in Fig.~\ref{fig:age_xfe_xh_nxh}f at around $9\,\mathrm{Gyr}$. At face value, these differences would exclude both the age-[Mg/Fe] and age-[Ba/Fe] plane as useful diagnostics, whereas all other elements show offsets in the age-[X/Fe] plane for at least the youngest ($8-11\,\mathrm{Gyr}$) accreted stars.

Moving towards the age-[X/H] plane in the middle row of Fig.~\ref{fig:age_xfe_xh_nxh}, we notice a very similar qualitative behaviour of all elements. However, we note that the separation of accreted and in-situ sequences in this plane is roughly five times larger, that is, around $0.5\,\mathrm{dex}$, than in the age-[X/Fe] plane. Finally, we consider the age-$N_\mathrm{X/H}$ plane in the bottom rows of Fig.~\ref{fig:age_xfe_xh_nxh}, where we again see a qualitatively similar trend. When focusing on the accreted sequence, we note a roughly linear enhancement of $N_\mathrm{X/H}$ for all elements except for Ba, which shows an increasingly enhanced curve towards younger ages, in line with an expected increase of Ba-production by AGB stars \citep{Karakas2016}. The difference between sequences starts very small at oldest ages, but then increases to $\Delta N_\mathrm{X/H} \sim 0.7$ around $8\,\mathrm{Gyr}$. While it is not the focus of this particular work, we want to note the increase in scatter of the in-situ age-abundance sequence, most notably in $N_\mathrm{X/H}$, for in-situ stars younger than the oldest accreted stars. This pattern has already been recognised in previous work on NIHAO-UHD simulations \citep[e.g.][]{Lu2022} and was connected with the time of disc formation for several Milky Way analogues. While \citet{Lu2022} identified the strong correlation with disc formation time (see their red lines in their Fig.~1), they also noted a correlation with major mergers \citep[][see their orange lines in Fig.~1]{Lu2022}. The analysis of different simulation snapshots is beyond the scope of this work. But for completeness we want to note the strikingly coinciding times of the onset of scatter in the in-situ sequence and end of star formation in the accreted sequence for the NIHAO-UHD Milky Way analogue of our study, \texttt{g8.26e11} which signifies the point in time of accreting the dwarf galaxy progenitor onto the Milky Way.

\begin{figure*}
	\includegraphics[width=\textwidth]{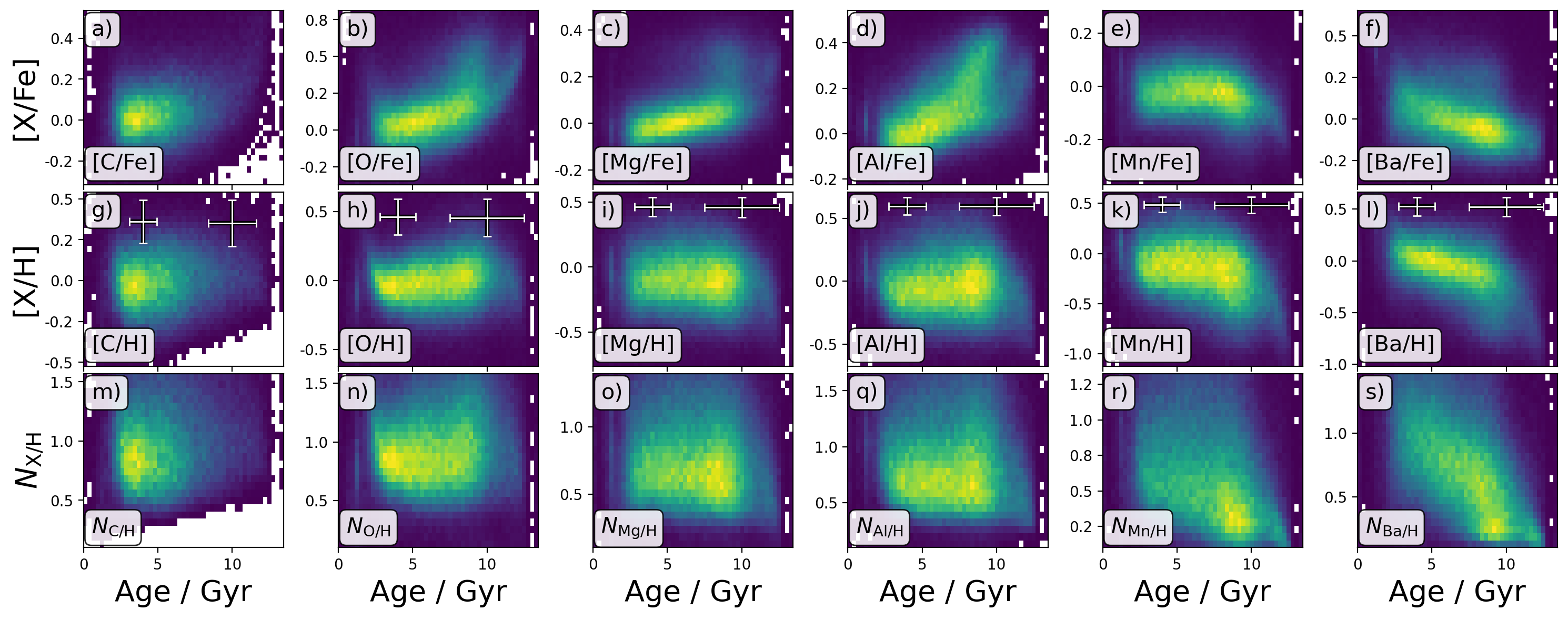}
    \caption{
    \textbf{Age-Abundance trends of the Milky Way Solar neighbourhood as observed by GALAH DR3.}.
    \textbf{Columns} show the different the elements C, O, Mg, Al, Mn, and Ba (from left column to right column), respectively.
    \textbf{Rows} show different notations of abundances, that is, logarithmic abundance {[X/Fe]} relative to iron (top row), logarithmic abundance {[X/H]} relative to hydrogen (middle row), and linear number density ratio {$N_\mathrm{X/H}$} (bottom row). Error bars in middle row indicate median uncertainties for stars below and above $7\,\mathrm{Gyr}$.
    }
    \label{fig:age_xfe_xh_nxh_galah}
\end{figure*}

\begin{figure*}
	\includegraphics[width=\textwidth]{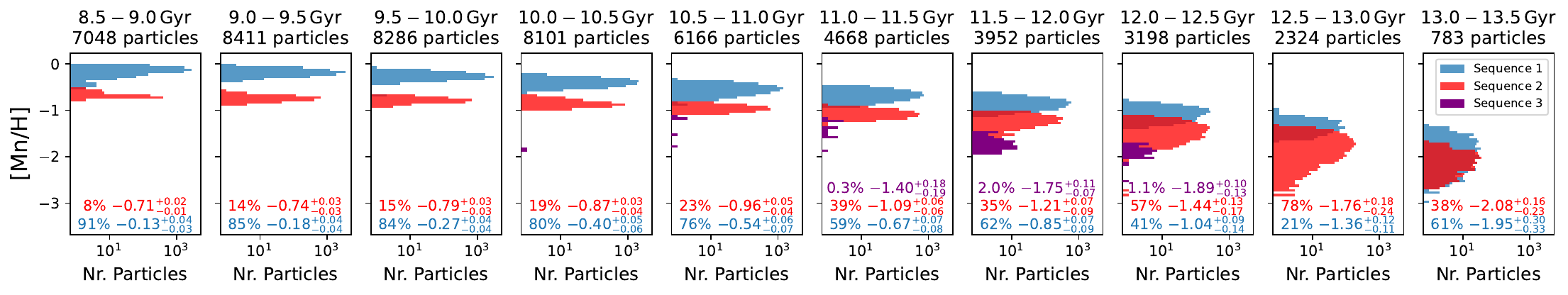}
    \caption{
    \textbf{Histograms of {[Mn/H]} in $0.5\,\mathrm{Gyr}$ age bins for the three age-abundance sequences (Eqs.~\ref{eq:sequence1}-\ref{eq:sequence3}).} Colors are the same as in Fig.~\ref{fig:three_sequences_traced}. Panel titles show age bins and total number of particles within each bin. Inset text with same colors show contribution (in percent) of each sequence to the age bins as well as 16/50/84th percentile. Values are only calculated if more than 100 particles of sequence are within a given age bin.
    }
    \label{fig:histograms_xh_in_age_bins_Mn}
\end{figure*}

Returning to the main focus of this work, that is, to better identify accreted stars in observations, we have created Fig.~\ref{fig:age_xfe_xh_nxh_galah} with observational data from the GALAH survey (Sec.~\ref{sec:obs_data}). While we see similarities between observations and simulated age-abundance trends for several elements, other elements show different behaviour. In particular, the measurements of C in GALAH are less numerous for older stars. This is expected based on the difficulty of measuring C from the atomic CI line. We will therefore refrain from a comparison for this element. For O, Mg, and Al, we see an increase of [X/Fe] towards old ages, For Al (Fig.~\ref{fig:age_xfe_xh_nxh_galah}d), this is contrary to the trend of the simulations (Fig.~\ref{fig:age_xfe_xh_nxh}d). The plots with respect to [X/H] are rather flat and inconclusive. For Mn, we see a decrease of both [Mn/Fe] and [Mn/H] towards old ages, in agreement with the simulation. For Ba, we see a decrease of both [Ba/Fe] and [Ba/H] towards old ages - in agreement with the simulations, but even more interestingly the hint of an overdensity of high $\mathrm{[Ba/Fe]} \sim 0.2$ and low $\mathrm{[Ba/H]} \sim -0.6$ around $9\,\mathrm{Gyr}$. This unique behavior is also seen in the simulations - and also only for Mn. When looking at the age-$N_\mathrm{X/H}$ plots in the lower row of Fig.~\ref{fig:age_xfe_xh_nxh_galah}, we identify the aforementioned overdensities again for $N_\mathrm{Ba/H}$ at $9\,\mathrm{Gyr}$, but also $N_\mathrm{Mn/H}$ (Fig.~\ref{fig:age_xfe_xh_nxh_galah}r), whereas the other elements C, O, Mg, and Al do not reach the regime of $N_\mathrm{X/H} < 0.4$ in larger numbers.

\textbf{Key Takeaway:} Across the different age bins, we find Gaussian distributions of abundances in the {[X/Fe]} and {[X/H]} space. At face value, the abundance vs age plots are separated most clearly when the abundance are noted relative to H rather than Fe. Establishing the relationships relative to the number density $N_{X/H}$ should also be explored in future work as they bear the potential to allow the tracing of linear and non-linear contributions through chemical yields.

\subsubsection{Quantifying the difference of age-{$\mathrm{[X/H]}$} sequences}

In this section, we are concerned with the significance of separation between the in-situ and accreted sequences in the simulations - for now in the idealised case without observational noise (we will study the latter in Sec.~\ref{sec:noise_influence}).

\begin{figure*}
	\includegraphics[width=\textwidth]{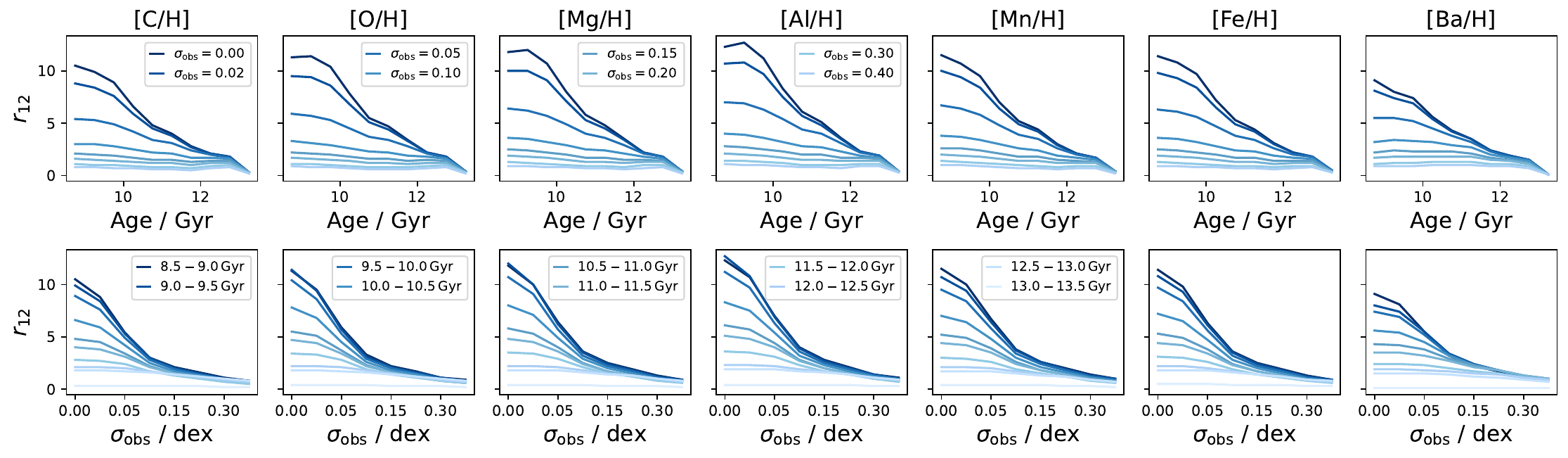}
    \caption{
    \textbf{Separation significance $r_{12}$ between Sequences 1 and 2 across different elements (panels) as a function of age bins (top panels) and added observational noise (bottom panels).} Lines are calculated for added observational noise in the top panels and different age bins in the bottom panels, respectively, as indicated in the legend.}
    \label{fig:r12_for_sigmaobs}
\end{figure*}

For this purpose, we trace the distribution of abundances [X/H] for the three sequences defined by Eqs.~\ref{eq:sequence1}-\ref{eq:sequence3} in histograms across different age bins. In Fig.~\ref{fig:histograms_xh_in_age_bins_Mn}, we show the histograms for the different sequences in age bins of $0.5\,\mathrm{Gyr}$ above $8.5\,\mathrm{Gyr}$, the youngest age of stars in Sequence 2. The panels of each figure include further useful information. Following the color-scheme from Fig.~\ref{fig:three_sequences_traced}, they state the relative contribution of each sequence to the age bin in percent, followed by the median abundance as well as the difference to the $16^\mathrm{th}$ and $84^\mathrm{th}$ percentile. We note firstly that the contribution of accreted stars is increasing from $8\%$ for $8.5-9.0\,\mathrm{Gyr}$ to $78\%$ for $12.5-13.0\,\mathrm{Gyr}$. In addition to the decrease of overall [Mn/H] towards older ages, we note that the scatter of both distributions increases from the youngest stars ($\sim 0.02-0.05\,\mathrm{dex}$) to the oldest ($\sim 0.1-0.4\,\mathrm{dex}$), while the distributions remain in a Gaussian shape for the most part, with larger tails towards lower [Mn/H] only noticeable for the oldest stars ($> 12.5\,\mathrm{Gyr}$).

\begin{figure*}
	\includegraphics[width=0.87\textwidth]{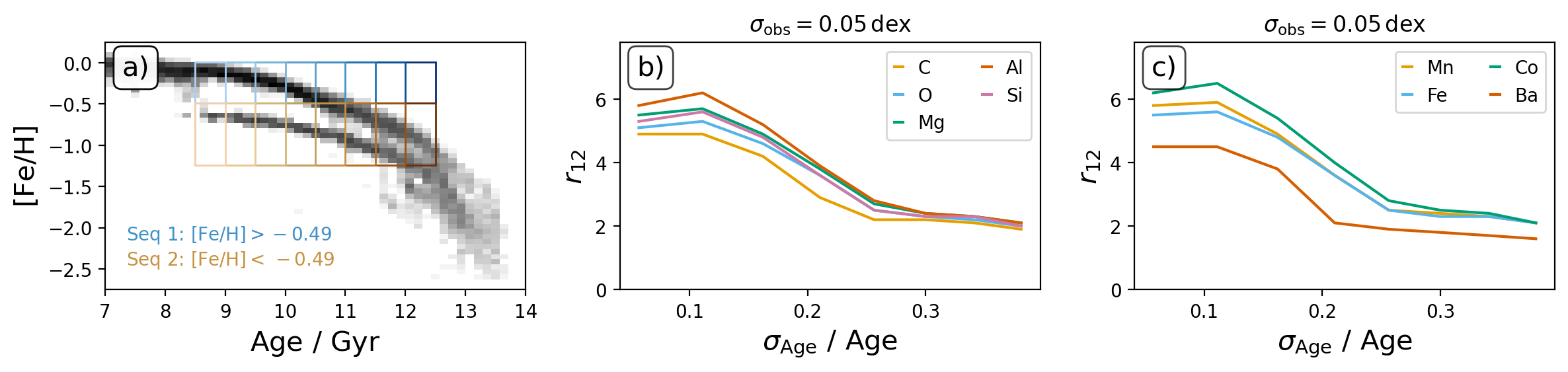}
    \caption{
    \textbf{Explanation of our gedankenexperiment (panel a) to quantify the remaining separation significance $r_{12}$ for larger age uncertainties for different elements.}
    \textbf{Panel a)} shows the age-{[Fe/H]} plane as example from which we select the two sequences with increasingly larger age bins from $8.5-9\,\mathrm{Gyr}$ to $8.5-12.5\,\mathrm{Gyr}$. \textbf{Panels b) and c)} show the resulting separation significance $r_{12}$ for different elements when selecting sequences based on the increasing bins (recalculated as relative bin sizes that can be understood as relative age uncertainties).
    }
    \label{fig:age_sigma_vs_r}
\end{figure*}

\begin{figure*}
	\includegraphics[width=\textwidth]{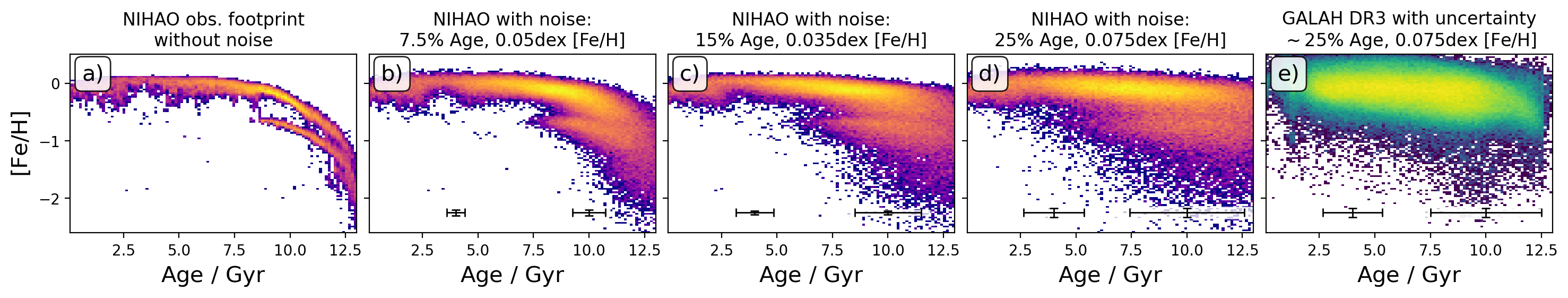}
    \caption{
    \textbf{Comparison of age-metallicity relation for different scatter realisations for NIHAO simulations.}
    \textbf{Panel a)} shows NIHAO without scatter, \textbf{panels b-e} show realisations of the same plot with increasingly more scatter sampling representative of uncertainties quoted in pioneering surveys (see text for more discussion). \textbf{Panel f} shows the GALAH DR3 age-metallicity relationship for comparison.
    }
    \label{fig:nihao_with_scatter}
\end{figure*}

To determine the significance of separation, we are inspired by the previous study by \citet{Lindegren2013}, who computed a separation significance $r$ based on the differences of mean abundances weighted by the combined standard deviation of abundances. Having convinced ourselves that the distributions are Gaussian in the previous section, we now expand their formulation by adding observational Gaussian noise as another additive to the weights and thus arrive at the formulation
\begin{equation} \label{eq:r_value}
r_{12} (\sigma_{obs}) = \frac{|\mu_1 - \mu_2|}{\sqrt{\sigma_1^2 + \sigma_2^2 + 2\cdot \sigma_{obs}^2}} ~ \xrightarrow[\text{for Sim.}]{\sigma_\text{obs}\,=\,0} ~ r_{12} = \frac{|\mu_1 - \mu_2|}{\sqrt{\sigma_1^2 + \sigma_2^2}}
\end{equation}

Assuming no observational noise, we calculate $r_{12}$ based only on the means and standard deviations for the age bins and elements {[X/H]} listed in Tab.~\ref{tab:tabular_separation_r12} for the sequences defined in Eqs.~\ref{eq:sequence1}-Eqs.~\ref{eq:sequence3}. For completeness, we also list the properties of Sequence 3, when more than 100 particles are present.

Table~\ref{tab:tabular_separation_r12} shows that the most significant separations $r_{12} > 10$ in the age-[X/H] space are expected among the youngest accreted stars. The separation significance decreases in order P (12.7), Al (12.7), Co (12.6), V (12.2), Mg (12.0), Cr (11.8), Mn (11.5), Si (11.5), Fe (11.4), Ne (11.4), O (11.4), S (11.1), C (10.5), N (10.3), and Ba (9.1)%
.

\textbf{Key Takeaway:} The intrinsic scatter in {[X/H]} for both in-situ and accreted sequences is larger ($\sim 0.1-0.4\,\mathrm{dex}$) for older stars ($> 12\,\mathrm{Gyr}$) as a result of larger change in this logarithmic scale when compared to stars born at the younger age of $8.5-11.0\,\mathrm{Gyr}$ ($\sim 0.02-0.05\,\mathrm{dex}$). This would make the younger accreted stars a better candidate to trace the chemical enrichment history and infer progenitor masses than old stars. While most elements, with the exception of the less suitable Ba, showed a similarly large separation significance, the odd-Z elements P and Al would be slightly more preferable than the iron-peak and alpha-process elements, which all outperform the lightest simulated elements C, N, O, and Ne.

\subsubsection{A more realistic picture: The influence of observational noise on age-{$\mathrm{[X/H]}$} sequence} \label{sec:noise_influence}

Having determined the most suitable plots for identifying accreted populations in simulations, we now consider this problem from a more realistic perspective and quantitatively determine how much noise we could allow for both abundance and age estimates to still be able to separate the sequences.

To get an intuition for the influence of measurement noise on the separation significance of abundances, we test different ranges of abundance uncertainties of $\sigma_\mathrm{obs}$ as input for Eq.~\ref{eq:r_value}. The resulting values are visualised in Fig.~\ref{fig:r12_for_sigmaobs} and representative values listed in Tab.~\ref{tab:tabular_separation_r12_noise}, which also includes median uncertainties of the measurements for GALAH DR3 \citep{Buder2021}, ranging from $\sigma_\mathrm{GALAH} = 0.06$ for [Si/H] to $\sigma_\mathrm{GALAH} = 0.13$ for [O/H] and [V/H].

We expect the sequences in age-[Fe/H] will be significantly more difficult to distinguish in older stars with corresponding larger uncertainties. To get a better intuition, at which age uncertainties we gain or loose more diagnostic power, we are performing a simple experiment. Assume that we know that our best chances to distinguish the sequences is to look at the youngest accreted stars, that is, those within our data set closest to $8.5\,\mathrm{Gyr}$. Using the smallest reasonable age bin ($8.5-9\,\mathrm{Gyr}$), we estimate a reasonable separation of Sequence 2 to 1 by estimating the mean and standard deviation of the sequence and adding a reasonable observational noise of $\sigma_\mathrm{obs} = 0.05\,\mathrm{dex}$ to it. This allows us to discern Sequence 1 from 2 with a rather simple cut of stars being above or below $\mu_2 + 3 \sqrt{\sigma_2^2 + \sigma_\mathrm{obs}^2}$, as shown in Fig.~\ref{fig:age_sigma_vs_r}a for $\mathrm{[Fe/H]} \lessgtr -0.49$. Using this abundance limit as the boundary, we can then test the distributions of Sequence 1 and 2 for increasingly large age bins of $0.5..(0.5)..4.0\,\mathrm{Gyr}$, as visualised with the selection boxes in Fig.~\ref{fig:age_sigma_vs_r}a. Performing this experiment for the various elements, we reach the distributions of $r_{12}$ in Figs.~\ref{fig:age_sigma_vs_r}a and \ref{fig:age_sigma_vs_r}b with increasing age bins size (which we equate with twice the age uncertainty).

Because we are working within the age range of $8.5-12.5\,\mathrm{Gyr}$, this means that an age bin of $2\,\mathrm{Gyr}$ would be equivalent to an age uncertainty of just over 20\%. We can see from the plot that we require an uncertainty of less than 20\% to achieve the highest separation significance $r$-value. While being an insightful gedankenexperiment, our tests on the influence of abundance and age uncertainties, however, neglect an important fact. That is, that both uncertainties are present at the same time in observations in complex ways and a selection like ours with prior knowledge of the sequence positions is not possible for the Milky Way.

Guided by our previous results on abundance and age uncertainties, and with the actual uncertainties of GALAH DR3 in mind, we have added both noises to the properties of the simulated galaxy and replicated the age-[Fe/H] in Fig.~\ref{fig:nihao_with_scatter}. This idea was also neatly implemented by \citet[][see their Fig.~12]{Renaud2021} to create more realistic noisy data for the \textsc{VINTERGATAN} simulations. Our figure shows the extremes of no noise on the far left (Fig.~\ref{fig:nihao_with_scatter}a) and the actual observed Milky Way age-[Fe/H] relation in the Solar neighbourhood on the far right (Fig.~\ref{fig:nihao_with_scatter}e with an average age uncertainty of $\sim 25\%$ and [Fe/H] uncertainty of $0.075\,\mathrm{dex}$). In between, we show different realisations of added Gaussian noise to the simulations. The chosen noise levels mimic those reported by pioneering studies, such as the one by \citet{Xiang2022} with a reported median uncertainty of $7.5\%$ for age and $0.05\,\mathrm{dex}$ for [Fe/H] (see Fig.~\ref{fig:nihao_with_scatter}b), the series of studies by \citet{Nissen2010} and \citet{Schuster2012} reporting uncertainties of $0.03-0.04\,\mathrm{dex}$ for [Fe/H] and a median of $15\%$ for age (Fig.~\ref{fig:nihao_with_scatter}d), as well as the already mentioned GALAH DR3 age-[Fe/H] relation (compare Figs.~\ref{fig:nihao_with_scatter}d~and~\ref{fig:nihao_with_scatter}e) with uncertainties of around $25\%$ for age and $0.075\,\mathrm{dex}$ for [Fe/H]. 

We want to stress here, that we assume that the uncertainties of metal-rich and metal-poor as well as young and old stars are relatively the same. In reality, this is not the case, because the measurements of old and metal-poor stars are less certain \citep{Frebel2015} when compared to stars with similar ages and metal-content than our Sun. The estimation of uncertainties - most importantly accuracy - is complicated. Most studies and surveys thus only report precision uncertainties, which neglects uncertainties of stellar evolutionary models \citep{Soderblom2010} just as much as uncertainties of abundance inference from stellar spectra, for example with approximations made in the modelling of atmospheres with uncertain line transitions \citep{Mihalas1973, Asplund2005}, among other factors \citep{Jofre2017, Jofre2018, Nissen2018}.

Setting these concerns aside when looking at Fig.~\ref{fig:nihao_with_scatter}, we can easily see that a larger noise in either direction blur out the clearly separated sequences of the noiseless simulation. While we can still see the separation of sequences in Figs.~\ref{fig:nihao_with_scatter}b~and~\ref{fig:nihao_with_scatter}c for the youngest accreted stars, the sequences overlap too much in Fig.~\ref{fig:nihao_with_scatter}d to be able to tell them apart by eye. We would therefore not expect to be able to tell apart a similarly separated sequence in the observed age-[Fe/H] of GALAH DR3 (Fig.~\ref{fig:nihao_with_scatter}e) in this particular plane. It is therefore not surprising, that the now established substructures of the stellar disk and halo in the solar vicinity of the Milky Way were only seen by \citet[][see their Fig.~4]{Sahlholdt2022} and later by \citet{Xiang2022} who each looked at main-sequence turnoff and subgiant stars, for which observable properties change more significantly as a function of stellar age. In particular, \citet{Xiang2022} were able to resolve a ``Z-shaped structure'' in the age-[Fe/H] relation of stars with lower angular momentum (see their Fig.~2). We note that the third sequence (bottom of Fig.~\ref{fig:three_sequences_traced}) would not be detectable in this plane even with the currently lowest uncertainties.

For completeness, we also want to note that the simulation at similar noise levels to GALAH DR3 shows a much tighter disk sequence than the actual Milky Way (compare the top of Figs.~\ref{fig:nihao_with_scatter}d~and~\ref{fig:nihao_with_scatter}e). This abundance spread would only be replicated if we were to add a noise of $\sigma_\mathrm{[Fe/H]}> 0.15\,\mathrm{dex}$ to the simulation. This either suggests that the well-documented radial migration \citep[e.g.][]{Haywood2008b, Frankel2018} of more metal-rich stars from the inner Galaxy and more metal-poor stars from the outer Galaxy to the Solar vicinity is not happening at the same strength in the Milky Way analogue or that turbulent metal mixing is slightly underestimated in the NIHAO simulations \citep[see also discussion in][]{Buck2021}.

\textbf{Key Takeaway:} Testing the uncertainties in the age-abundance planes independently and in concert, we find critical uncertainties of $\sigma_{[X/H]} = 0.15\,\mathrm{dex}$ and $\sigma_\mathrm{age} = 20\%$ above which we loose the ability to tell apart accreted and in-situ sequences with certainty. Adding noise at the level of current pioneering studies to the simulations confirms the qualitative impression of a ``Z-shaped structure'' \citep{Xiang2022} in the noisy age-[Fe/H] plane for two underlying sequences (Fig.~\ref{fig:nihao_with_scatter}b).

\section{Discussion} \label{sec:discussion}

The initial motivation of this research was to create a variety of abundance planes and compare them using both the simulated and observed data to determine the best abundance-abundance plots for identifying substructure that could be representative of an accreted population of stars. In this section, we therefore discuss not only the question which information reveals accreted structures best (Sec.~\ref{sec:discussion_best_information}), but also take a closer look at the age-[Fe/H] relation around the time of the last merger and discuss implications of the found patterns (Sec.~\ref{sec:discussion_zoom_at_mergertime}). At the end of our discussion, we touch on how comparable the simulation at hand is to the Milky Way - and what that means for the findings of our comparison; also in the context of galaxy evolution (Sec.~\ref{sec:discussion_how_comparable_are_nihao_and_milkyway}).

\subsection{Which information reveals accreted structures best?} \label{sec:discussion_best_information}

In an evolving galaxy, the kinematic and dynamic information of stars that were born together is expected to change. This memory may not be completely lost throughout a galaxy's vivid evolution. This is exemplified by the many recovered structures \citep{Deason2024} and even metallicity gradients \citep{Khoperskov2023d} in velocity and energy-angular momentum space. Nonetheless, orbits are certainly expected to be blurred throughout evolution. Even within our simulated data, we see significant overlap in the kinematic characteristics of the three separate sequences (Fig.~\ref{fig:three_sequences_traced}). Depending on the size and time of mergers there could be significantly more overlap between accreted and in-situ populations.

Chemical compositions and ages - although harder to infer - are expected to be more conserved in stellar atmospheres than dynamic properties, a paradigm of the idea of \textit{chemical tagging} \citep{FreemanBlandHawthorn2002}. Our analyses, in agreement with previous analyses \citep[e.g.][]{Khoperskov2023c, Rey2023}, however found that while stars that were born together may share similar chemistry, this chemistry does not have to be significantly different from other birth places for the light elements inspected in this study. This problem brings the study by \citet{Ness2018} to mind, who found it difficult to tell apart stars of the stellar disk with light element abundances for $0.3\%$ of their inspected field star pairs. Extending this study to heavy elements with GALAH, \citep{Manea2023} found that heavy elements help to decrease such doppelgänger rates by an additional factor of 6. The simulation of this study does only follow up Ba with the currently limited understanding of neutron-capture enhancement. Although we have highlighted the interesting observational pattern of age-[Ba/H] and age-$N_\mathrm{Ba/H}$ earlier (see Sec.~\ref{sec:age_xfe_xh_nxh} and the discussion of Figs.~\ref{fig:age_xfe_xh_nxh_galah}f, \ref{fig:age_xfe_xh_nxh_galah}l, and \ref{fig:age_xfe_xh_nxh_galah}s), this avenue needs to be explored in future studies - both observationally and through chemical evolution modelling.

In some cases the chemical information of some structures (e.g. Sequence 3) is quite compact in abundance-abundance space (see Fig.~\ref{fig:three_sequences_traced}n for [Fe/H] vs. [Si/Fe]). A careful selection and combination may still provide a useful avenue to trace back such substructure that has lost its kinematic memory, even though not all chemical elements may provide useful insight (see e.g. for Al in Fig.~\ref{fig:three_sequences_traced}o).

Again, it is important to note that the simulated galaxy may have different characteristics to our own Galaxy. As discussed, there are two clear streams that have been attributed as accreted and in situ, based on the assumption that the accreted stars would have been born in a smaller galaxy with less massive stars and therefore lower element abundances.

\begin{figure*}
	\includegraphics[width=0.9\textwidth]{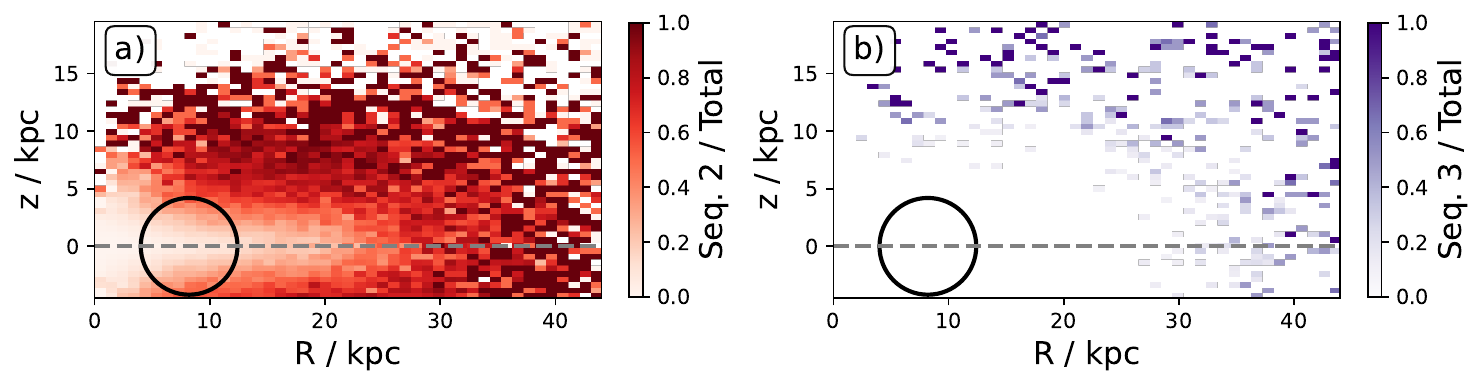}
    \caption{
    \textbf{Relative contributions of Sequence 2 (panel a) and Sequence 3 (panel b) across the galactic area $R$ vs. $z$.} The observational footprint within the Milky Way is indicated with a black circle and the galactic plane with a grey dashed line. The larger coverage of $R$ vs. $z$ with bin sizes $1$ and $0.5\,\mathrm{kpc}$, respectively, is inspired by the extragalactic study by \citet[][their Fig.~16]{Martig2021}.
    }
    \label{fig:accretion_rate}
\end{figure*}

\begin{figure*}
	\includegraphics[width=\textwidth]{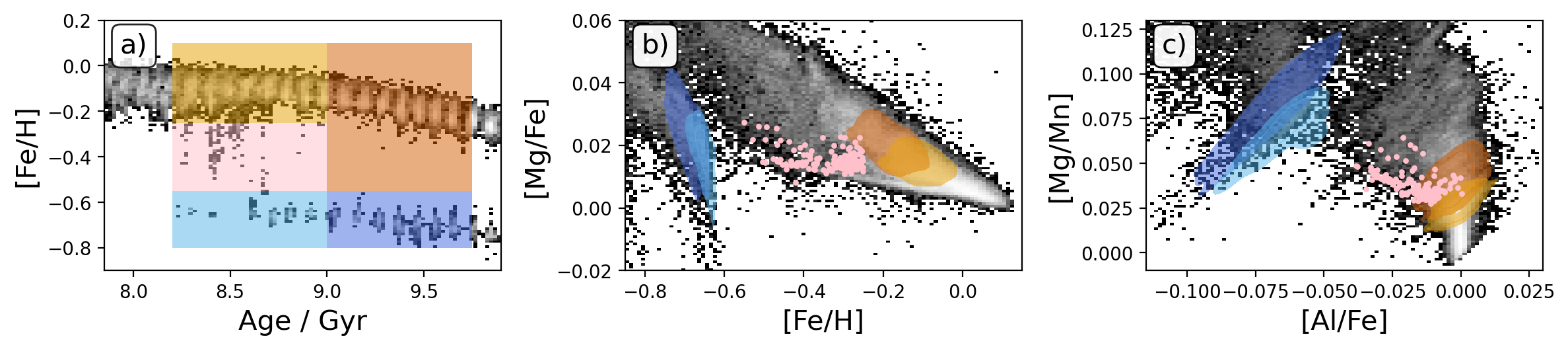}
\caption{
    \textbf{Tracing the chemistry of particles in different regions of the age-[Fe/H] relation for the whole Milky Way analogue.}
    \textbf{Panel a)} shows the age-[Fe/H] relation as grey density plot with the five rectangles indicating the particles that are traced.
    \textbf{Panel b)} shows the [Fe/H] vs. [Mg/Fe] plane with the density contours (or scatter points in the case of the intermediate pink region) of the different traced particles.
    \textbf{Panel c)} shows the [Al/Fe] vs. [Mg/Mn] plane for the same particles.
    }    \label{fig:tracing_amr_regions_footprint}
\end{figure*}

While the jury on the best information for identifying is still out, this study yet again underlines that the survey of the whole galaxy is important to identify smaller accretion events. While for example a non-negligible amount of stars of the accreted Sequence 2 can be found in the observable footprint (see Fig.~\ref{fig:accretion_rate}a), we find stars of Sequence 3 only far outside of the GALAH-like footprint at galactic heights of $\vert z \vert > 10\,\mathrm{kpc}$ (see Fig.~\ref{fig:accretion_rate}b). This area is being probed by the H3 Survey \citep{Conroy2019} in the actual Milky Way and has indeed found a significant amount of substructure, with more findings expected by the upcoming 4MOST halo surveys \citep{4MOST_HR_Halo, 4MOST_LR_Halo} as well as the 4MOST dwarf galaxies and streams survey \textit{4DWARFS} \citep{4DWARFS2023}.

\subsection{A closer look at age-[Fe/H] relation around the last major merger} \label{sec:discussion_zoom_at_mergertime}

While stellar mass estimates can help us to quantify the mass ratios of previous mergers, we ultimately want to probe not just the presence and correlation but the causality of major mergers and chemical thin/thick disk bimodality, among other quantities. To this end, we are thus focusing on the age-[Fe/H] relation around the time of the last major merger in this section. While this is important both in terms of stellar particles and gas particles \citep[see e.g. the study by][]{Buck2023}, this study focuses on the stellar particles as the surviving and observable tracers throughout galactic evolution.

In Fig.~\ref{fig:tracing_amr_regions_footprint}, we are thus zooming into the age-[Fe/H] relation around $8.6\pm0.4\,\mathrm{Gyr}$, the rough\footnote{Quantifying the beginning and end of a merger process with all its intricate dynamical and chemical interactions is difficult. However, our rough estimate of $8.6\pm0.4\,\mathrm{Gyr}$ is almost twice as precise as current best age precision uncertainties.} time of the last merger. Around this time, we notice not only the stop of star formation in the accreted sequence (see light blue region in Fig.~\ref{fig:tracing_amr_regions_footprint}a), but also the appearance of several particles with iron abundance between the in-situ and accreted sequence (see pink region in Fig.~\ref{fig:tracing_amr_regions_footprint}a). We thus trace several properties of the in-situ (dark and light orange) as well as accreted sequence (dark and light blue) before and throughout the merger, respectively. In particular, we notice a continuing increase of [Fe/H] between $9-9.5\,\mathrm{Gyr}$ to $8-8.5\,\mathrm{Gyr}$ for both sequences. For the in-situ sequence, we notice an increase from $-0.165 \pm 0.038$%
 to $-0.113 \pm 0.056$%
. In particular, we find an increase of the [Fe/H] spread between $9-9.5\,\mathrm{Gyr}$ to $8-8.5\,\mathrm{Gyr}$ from 
0.038%
 to 0.056%
, that is an almost 50\% increase. When calculating the same quantities for the whole galaxy (Fig.~\ref{fig:tracing_amr_regions_whole}), we find a smaller increase of [Fe/H] spread from 0.065%
 to 0.076%
, that is only 17\%. For the accreted sequence, we find an increase from $-0.682 \pm 0.029$%
 to $-0.634 \pm 0.031$%
 with consistent scatter. We further find a decrease in the ratios of [Mg/Fe] and [Mg/Mn] in Figs.~\ref{fig:tracing_amr_regions_footprint}b and \ref{fig:tracing_amr_regions_footprint}c, whereas the ratio of [Al/Fe] stays rather constant for both sequences (Fig.~\ref{fig:tracing_amr_regions_footprint}c).

Most interesting to us seem the pink particles of Fig.~\ref{fig:tracing_amr_regions_footprint} with [Fe/H] between the in-situ and accreted sequences. These arise throughout the merger with ages of $8.6\pm0.4\,\mathrm{Gyr}$ and are also exhibiting intermediate abundances in [Mg/Fe], [Al/Fe], and [Mg/Mn] in Figs.~\ref{fig:tracing_amr_regions_footprint}b~and~\ref{fig:tracing_amr_regions_footprint}c. Such stars have previously been identified as stars that formed further outside of the Solar vicinity  with larger angular momenta \citep[e.g.][]{Haywood2008b, Haywood2013, Buder2019}. We too find them to have larger angular momenta of $L_Z=1340_{-490}^{680}\,\mathrm{kpc\,km\,s^{-1}}$%
 when compared to $L_Z=1290_{-480}^{600}\,\mathrm{kpc\,km\,s^{-1}}$%
 for stars of the in-situ disk at the same time and even higher than the accreted stars with same ages ($L_Z=1140_{-600}^{700}\,\mathrm{kpc\,km\,s^{-1}}$%
). The relative contribution of stars with these properties is, however, relative little with a stellar mass of only $0.1\times10^{8}\,\mathrm{M_\odot}$%
 compared to the amount of stars born in the dark and bright orange in-situ regions ($6.7\times10^{8}\,\mathrm{M_\odot}$%
 and $6.1\times10^{8}\,\mathrm{M_\odot}$%
) and even the accreted stars with their decreasing star formation between dark and bright blue regions ($1.2\times10^{8}\,\mathrm{M_\odot}$%
 and $0.4\times10^{8}\,\mathrm{M_\odot}$%
). Even when extending the analysis towards younger in-situ stars with lower [Fe/H] (see Fig.~\ref{fig:tracing_amr_regions_disk}) we continue to find relative low numbers of these stars, whereas they seem to be more numerous in the Milky Way's Solar vicinity \citep{Haywood2013, Hayden2015, Buder2019}, thus raising important question of the similarity of Milky Way and simulated analogues which we discuss in the next subsection.

\subsection{How comparable is the NIHAO Milky Way analogue to the Galaxy?} \label{sec:discussion_how_comparable_are_nihao_and_milkyway}

Throughout this study we have tried to quantify the usefulness of specific diagnostic plots based on noiseless simulations of a Milky Way analogue. In this section of the discussion, we are now concerned with the question, how similar the Milky Way analogue is to the Milky Way and to which extend our conclusions can thus be applied to the Milky Way.

\paragraph*{Chemical comparability:} As we acknowledged throughout this study (see e.g. Sec.~\ref{sec:sim_data}), the chemical evolution of Milky Way analogue simulations is neither fully accurate nor consistent with the Milky Way. While the basic properties of the galaxy at hand agree with the Milky Way within the order of magnitude, we acknowledge that a different merger history may lead to different chemical compositions. More of concern for our study is the inaccuracy and incompleteness of chemical enrichment parameters, such as chemical yields. While we can overcome potential relative inaccuracies in element production by up- or down-scaling the total number densities of elements (which in a logarithmic abundance notation leads to simple shifts, as tabulated in Tab.~\ref{tab:shift_4to5gyr}), we cannot correct missing channels - for example for Ba or other heavy elements. This certainly means that the chemical evolution and predicted sequence separations have to be taken with a grain of salt and can only be used as indicators.

\paragraph*{Merger mass ratio:} For the Milky Way, different estimates of the last major merger have been estimated through different comparisons. These include the stellar mass estimates of $5\times10^{8}\,\mathrm{M_\odot}$ based on simulations and dynamics \citep{Naidu2021} in agreement with findings by \citet{Helmi2018} based on chemistry and simulations ($6\times 10^{8}\,\mathrm{M_\odot}$), but below the estimates of \citet{Feuillet2020} ($7-70\times10^{8}\,\mathrm{M_\odot}$). Halo mass ratios have been estimated between 1:2.5 \citep{Naidu2020} and 1:4 \citep{Helmi2018}, when also taking into account dark matter masses \citep[e.g. $2\times10^{11}\,\mathrm{M_\odot}$ for][]{Naidu2020}.

\begin{figure}
	\includegraphics[width=\columnwidth]{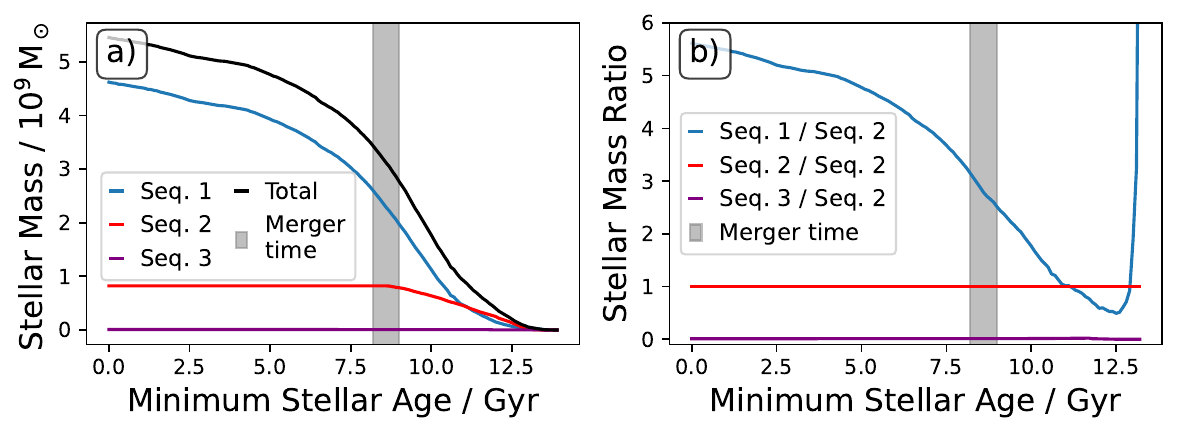}
    \caption{
    \textbf{Absolute (panel a) and relative (panel b) stellar mass contributions to the observed solar neighborhood as a function minimum stellar age.}
    \textbf{Panel a)} show the stellar mass ratio of the major accreted sequence to the in-situ sequence.
    }
    \label{fig:sequence_mass_ratios}
\end{figure}

When tracing the stellar masses of the in-situ and accreted sequences (Fig.~\ref{fig:sequence_mass_ratios}), we find a similar stellar mass ratio of 1:3.0%
 around $8.2-8.6\,\mathrm{Gyr}$ based on stellar masses of 0.8%
$\times10^9\,\mathrm{M_\odot}$ and 2.4%
$\times10^9\,\mathrm{M_\odot}$ for stars older than $8.6\,\mathrm{Gyr}$.

\paragraph*{The (missing?) separation of thin and thick disk:}
Contrary to the more distinct sequences that were found by \cite{Nissen2020, Sahlholdt2022, Xiang2022} for thin and thick disk stars in the age-[Fe/H] relation, we do not see a significant change along the age-[Fe/H] relation of in-situ component (as in thin vs. thick disk). Similarly, the [Fe/H] vs. [Mg/Fe] sequence does not present itself as the clear bimodal one seen in the Milky Way \citep{Hayden2015} and other NIHAO simulations \citep{Buck2020, Buck2023}. This has several reasons: a) the Milky Way analogue discussed in this study uses a different yield set \citep[\textsc{alt} as shown in][]{Buck2021} but traces many more elements. Comparing to Fig. 9 of \citet{Buck2021} we can see that this yield set produces a tighter sequence not quite resembling the observed one. b) the particle mass of the simulation is $10^5\,\mathrm{M_\odot}$ for star particles which for numerical reasons diminishes metal mixing and a more clearer separation of the sequences. Additionally this means we are not resolving all merging dwarf galaxies but only the most massive ones. Nevertheless, the approach taken in this study by only looking at differential analyses of abundance patterns are robust findings. The presence of a thick disk may not necessarily be seen as a clear separation in [Fe/H] vs. alpha-enhancement, despite manifesting itself in the kinematics \citep[see e.g.][their Fig.~12]{MCM2013}.

This Milky Way analogue did only undergo one significant merger around $8.5\,\mathrm{Gyr}$ ago. We find an absence of a clearly chemically distinct disk with a more or less significant interruption in star formation and change in later chemical evolution of disk stars. A similar finding was made by \citet{MCM2013} where in the absence of early-on massive mergers, their thick disk was kinematically less hot with vertical velocity dispersions smaller by a factor of $\sim 2$ when compared to the Milky Way. Extending their conclusion into the chemical dimension, our analysis of only one(!) Milky Way analogue could lead to a similar conclusion as theirs, that is, that ``\textit{the Milky Way thick disk is unlikely to have been formed through a quiescent disk evolution}'' \citep{MCM2013}.

\paragraph*{The (missing?) spread in the age-[Fe/H] relation of the young disk:}

In Sec.~\ref{sec:discussion_zoom_at_mergertime}, we pointed out the low spread of [Fe/H] among young disk stars. From 9.0-9.5, 8.5-9.0, and $8.0-8.5\,\mathrm{Gyr}$, we see a change of [Fe/H] from $-0.166 \pm 0.037$%
 to $-0.115 \pm 0.037$%
 to $-0.109 \pm 0.050$%
 in the simulation. This compares to a change of $-0.222 \pm 0.255$%
 to $-0.203 \pm 0.258$%
 to $-0.186 \pm 0.256$%
 for GALAH DR3, respectively\footnote{Here we report median values and half the difference between $16^\mathrm{th}$ and $84^\mathrm{th}$ percentiles. Assuming a Gaussian distribution, we would measure larger scatters with a qualitatively similar behaviour for the overall [Fe/H] values of $-0.165 \pm 0.038$%
,  $-0.114 \pm 0.043$%
, and  $-0.113 \pm 0.056$%
 for the simulation and $-0.251 \pm 0.292$%
,  $-0.225 \pm 0.281$%
, and $-0.205 \pm 0.271$%
 for GALAH DR3.}. This significantly lower spread of the simulation could either specifically point towards a different formation history of disk stars in our simulation because of a much lower radial migration than in the actual Milky Way  \citep{Minchev2010, Loebman2011, Frankel2018} or reflect a suppressed metal mixing in the simulations.

\paragraph*{Bottom line:}
We acknowledge that the discussion of this particular section is incomplete and more comparisons could be done. However, we believe we have established that in terms of accretion, the simulation at hand can provide useful insight and we therefore conclude that further discussion about the comparability of simulation and observation (such as gas masses or the presence and influence of a bar or spiral arms) are important next steps but beyond the scope of this work.
Such research should be performed also with other cosmological simulations, preferably with varied chemistry and formation histories. This would enable researchers to observe exactly how different variables in the simulation affect outcomes. Such comparisons would not only be insightful for the Milky Way, but also observations of other galaxies \citep[compare for example figures by][to our Fig.~\ref{fig:accretion_rate}]{Pinna2019, Pinna2019b, Martig2021} and the ongoing observational projects such as the large MUSE program GECKOS \citep{GECKOS2023}.

\section{Conclusions}
\label{sec:conc}

Both the research field of Galactic archaeology in particular and galaxy evolution as a whole aim to unravel how galaxies form and evolve across cosmic time. While the observation of galaxies across different redshifts can provide us with insights into the pathways of galaxy evolution, the Milky Way provides us with the unique resolved information at redshift zero. To link all of these different snapshots, we can use the numerous scenarios provided by simulations, which now also allow us to trace the chemical evolution across different elements based on our current best understanding of nucleosynthesis. Mergers of galaxies and accretion of smaller systems are common events in a Universe that forms hierarchically and we see evidence of such mergers especially at high redshift $z > 2$ (or more than $10\,\mathrm{Gyr}$ ago).

An outstanding question is, if mergers play a minor role or major role in shaping galaxies. Do they for example only drive the scatter in the relation of age (or redshift) with velocity dispersion and metallicity, or do they actually set these relations and influence the shape of galaxies through the triggering of disk and bar formation?

To get a quantitative handle on answering these questions, we need to trace back the merger history of the Milky Way as the one galaxy for which we can get a resolved understanding of evolution across time and space. This paper thus aims to find new insights from the cosmological \textsc{NIHAO} simulations on how we can identify the surviving stars of accreted galaxies.

\paragraph*{The major take-away points from this undertaking were:}
\begin{itemize}
    \item A first assessment of the common diagnostic plots of observations manifested the old credo that the absence of clear features or subpopulations in the data does not mean that they are not present - location and selection do matter (Sec.~\ref{sec:location}). While the assessment of chemical abundance planes of the Milky Way analogue for the whole galaxy is actually overshadowing the minor contribution of accretion, we confirmed the potential of age-abundance and abundance-abundance plots for smaller regions, both in similar footprint as current Milky Way observations (Fig.~\ref{fig:low_alpha_halo}) as well as extragalactic observations (Fig.~\ref{fig:accretion_rate}).
    \item In particular, when looking at abundance with respect to iron abundance (Sec.~\ref{sec:feh_xfe} and Fig.~\ref{fig:FeH_XFe}) as well as [Al/Fe] vs. [Mg/Mn] (Sec.~\ref{sec:alfe_mgmn} and Fig.~\ref{fig:NaFe_MgMn_selection_Age_FeH_dissection}) we confirmed the strong potential of chemistry to identify major substructures. At the same point, we realised that different accreted structures are likely occupying the same abundance space - as previously suggested by both \citet{Horta2021} and \citet{Rey2023}. Only smallest measurement uncertainties avoid the blurring of sequences (Fig.~\ref{fig:mgmn_alfe_with_noise}).
    \item When tracing three sequences in the age-[Fe/H]-plane, we confirmed that their significant overlap appeared when these were plotted in terms of dynamical and chemical properties (Fig.~\ref{fig:three_sequences_traced}). However, in the age-[Fe/H]-plane these sequences were clearly separated. We thus inspected the age-abundance relations in more detail (Sec.~\ref{sec:Age-abundance}). 
    \item Plotting elemental abundances against age, proved fruitful, with very distinct sequences of accreted and in situ populations appearing in the simulation data (Figs.~\ref{fig:age_xfe_xh_nxh} and \ref{fig:histograms_xh_in_age_bins_Mn}) with up to 12-sigma separation (Tab.~\ref{tab:tabular_separation_r12}) at the youngest ages.
    \item Such sequences are not as clear in the observational plots of GALAH DR3 (Fig.~\ref{fig:age_xfe_xh_nxh_galah}), as a result of massive uncertainties that come with the probabilistic nature of estimating stellar ages through isochrone fitting.
    \item When testing the significance of separations of sequences in the age-[X/H] plane, we see the largest separations for the youngest accreted stars, even at larger observational noise (Fig.~\ref{fig:r12_for_sigmaobs} and Tab.~\ref{tab:tabular_separation_r12_noise}). We find that we loose the ability to tell sequences apart beyond 2-sigma for observational noise above $0.15\,\mathrm{dex}$ (Fig.~\ref{fig:r12_for_sigmaobs}) and age uncertainties above $20\%$ (Fig.~\ref{fig:age_sigma_vs_r}).
    \item With observational data below these uncertainties, we expect to see accreted and in-situ sequences separated (Fig.~\ref{fig:nihao_with_scatter}), as has been the case in the work by \citet{Xiang2022}.
    \item In this work, we have only analysed one Milky Way analogue with one clear in-situ sequence and one major as well as one minor accreted sequence. For the major accreted sequence, we find a stellar mass ratio of 1:3.0%
 at the time of last star formation in the accreted sequence (Fig.~\ref{fig:sequence_mass_ratios}).
\end{itemize}

\paragraph*{Future Research:}
These insights provide clear avenues for following up - both in the observational data and other simulations beyond the one Milky Way analogue of this study:
\begin{itemize}
    \item We can use the observed chemical evolution in both in-situ and accreted components - formed in systems with different masses and stellar mass distributions - to adjust the chemical yields and thus chemical evolution in simulations (see Fig.~\ref{fig:FeH_XFe} for example for Ba as a neutron-capture element).
    \item Simulations are an excellent test bed to probe the efficiency of our observational tools to identify accreted stars. In particular, we plan to quantify the accuracy and completeness of previously used Gaussian Mixture Models  \citep{Das2020, Buder2022} with simulated [Al/Fe] vs. [Mg/Mn] plane in a follow-up paper.
    \item This study underscores the imperative for more precise and accurate age determinations. While better precision can be expected from additional information by asteroseismic missions \citep[see e.g.][]{Miglio2017,Mackereth2021}, better accuracy can only be provided by more advanced stellar evolutionary models \citep[see also][]{Kim2002, Schuster2012} as an undeniable bottleneck of Galactic archaeology.
    \item We need to quantify the separation of sequences in other NIHAO galaxies \citep[see e.g.][]{Lu2022, Buck2023} as well as other simulations, like \textsc{VINTERGATAN} \citep{Renaud2021, Renaud2021b, Agertz2021}, \textsc{HESTIA} \citep{Khoperskov2023, Khoperskov2023b, Khoperskov2023c} and if possible beyond just age-[Fe/H] \citep{Khoperskov2023c}. Constraining several of the simulation parameters, for example by up- or down-scaling merger significance to the point where they do or do not resemble our Milky Way properties is another exciting avenue along this line \cite{Rey2023}.
    \item Following up the correlation and causality of the spread in the age-[Fe/H] relation among stars born after the major merger (Fig.~\ref{fig:tracing_amr_regions_footprint}) is an exciting avenue of further research. It will be important to compare the stellar, gas, baryonic, and total masses, inclination angles, merger timescales, strength and emergence of bars and other properties of accreted galaxies in simulations to the impacts of massive mergers on the cold gas metallicity gradients \citep{Buck2023} as well as the correlation with observational plots that use age and/or chemistry. This will be complementary to the ongoing work in the kinematic/dynamic comparisons \citep[e.g.][]{Naidu2021, Khoperskov2023b} when reconstructing the complex pattern of accretion in the Milky Way \citep{Naidu2020}.
\end{itemize}

Fuelled by the revolutionary orbit information of the \textit{Gaia} satellite \citep{Brown2016, Brown2018, Brown2021, Vallenari2023} with its unprecedented astrometric \citep{Lindegren2016, Lindegren2018, Lindegren2021a} and spectroscopic information from both \textit{Gaia} \citep{Katz2019, Katz2022} as well as ground-based surveys \citep[e.g.][]{SDSSDR17, Buder2021, Zhao2012, Conroy2019}, we are currently in the discovery phase of substructure in our Milky Way from observational data; with new structures being identified in diverse and creative ways \citep[e.g.][]{Naidu2021, Malhan2024}. This surge in discovery is undeniably thrilling, yet it is worth to keep in mind that our bigger goal remains to quantify the role of accretion on shaping our Galaxy. A major question continues to be, if accretion was responsibly for the change in star formation efficiency around the end of thick disk and onset of thin disk formation \citep{Conroy2022} - or just happening at the same vivid time in our Galaxy's evolution \citep[see the reviews by][for more extensive discussions]{Helmi2020, Deason2024}. Our best way forward to answering this question is to connect the unique insights from resolved observations in the Milky Way with extragalactic observations and theoretical predictions from simulations \citep[see e.g.][]{GECKOS2023}.

\section*{Acknowledgements}

We thank Martin~P.~Rey for insightful discussions and valuable comments on the manuscript.

We acknowledge the traditional owners of the land on which the AAT and ANU stand, the Gamilaraay, the Ngunnawal and Ngambri people. We pay our respects to elders past, present, and emerging and are proud to continue their tradition of surveying the night sky in the Southern hemisphere.

This work was supported by the Australian Research Council Centre of Excellence for All Sky Astrophysics in 3 Dimensions (ASTRO 3D), through project number CE170100013. SB acknowledges support from the Australian Research Council under grant number DE240100150 which enabled SB to continue researching at the end of a fixed-term position and finalising this study. TB acknowledges funding from the Carl Zeiss Stiftung and support from the European Research Council under ERC-CoG grant CRAGSMAN-646955. We gratefully acknowledge the Gauss Centre for Supercomputing e.V. (\url{www.gaus s-centre.eu}) for funding this project by providing computing time on the GCS Supercomputer SuperMUC at Leibniz Supercomputing Centre (\url{www.lrz.de}). Simulations were partially computed with High Performance Computing resources at New York University, Abu Dhabi.

\section*{Facilities}

\textbf{AAT with 2df-HERMES at Siding Spring Observatory:} The GALAH Survey is based data acquired through the Australian Astronomical Observatory, under programs: A/2013B/13 (The GALAH pilot survey); A/2014A/25, A/2015A/19, A2017A/18 (The GALAH survey phase 1), A2018 A/18 (Open clusters with HERMES), A2019A/1 (Hierarchical star formation in Ori OB1), A2019A/15 (The GALAH survey phase 2), A/2015B/19, A/2016A/22, A/2016B/10, A/2017B/16, A/2018B/15 (The HERMES-TESS program), and A/2015A/3, A/2015B/1, A/2015B/19, A/2016A/22, A/2016B/12, A/2017A/14, (The HERMES K2-follow-up program). This paper includes data that has been provided by AAO Data Central (\url{datacentral.aao.gov.au}).

\textbf{\Gaia: } This work has made use of data from the European Space Agency (ESA) mission \Gaia (\url{http://www.cosmos.esa.int/gaia}), processed by the \Gaia Data Processing and Analysis Consortium (DPAC, \url{http://www.cosmos.esa.int/web/gaia/dpac/consortium}). Funding for the DPAC has been provided by national institutions, in particular the institutions participating in the \Gaia Multilateral Agreement. 

\textbf{Other facilities:} This publication makes use of data products from the Two Micron All Sky Survey \citep{Skrutskie2006} and the CDS VizieR catalogue access tool \citep{Vizier2000}.

\section*{Software}

The research for this publication was coded in \textsc{python} (version 3.7.4) and included its packages
\textsc{astropy} \citep[v. 3.2.2;][]{Robitaille2013,PriceWhelan2018},
\textsc{corner} \citep[v. 2.0.1;][]{corner},
\textsc{IPython} \citep[v. 7.8.0;][]{ipython},
\textsc{matplotlib} \citep[v. 3.1.3;][]{matplotlib},
\textsc{NumPy} \citep[v. 1.17.2;][]{numpy},
\textsc{pynbody} \citep[v. 1.1.0;][]{pynbody},
\textsc{scipy} \citep[version 1.3.1;][]{scipy},
\textsc{sklearn} \citep[v. 0.21.3;][]{scikit-learn},
We further made use of \textsc{topcat} \citep[version 4.7;][]{Taylor2005};

\section*{Data Availability}

All code and data to reproduce the analysis and figures can be accessed via \url{https://github.com/svenbuder/Accretion_Clues_ObsSim}.

The repository also includes the chemical and kinematic data of the simulated Milky Way analogue \texttt{g8.26e11} for last snapshot of the simulation for the observable footprint that were used for this study as well as the cleaned catalogue of GALAH DR3. All GALAH DR3 data is also published by \citet{Buder2021} and can be accessed publicly via \url{https://docs.datacentral.org.au/galah/dr3/overview/}. The full simulation data of \texttt{g8.26e11} can be obtained upon reasonable request from the authors. Currently the only limitation in making all data public is limited cloud space to host the data. We encourage interested readers to get in contact with the authors for full data access. Redshift zero snapshots from the original NIHAO-UHD simulations can be found here: \url{https://tobias-buck.de/\#sim_data}.

If you are using either of these data to follow up on this research, remember to give appropriate credit to the researchers who created and curated either data set, that is, at least to \citet{Buder2021, Buder2022} and \citet{Buck2020b, Buck2021}.


\bibliographystyle{mnras}
\bibliography{bib} 


\appendix

\section{Additional figures and tables} 

\begin{figure*}
	\includegraphics[width=\textwidth]{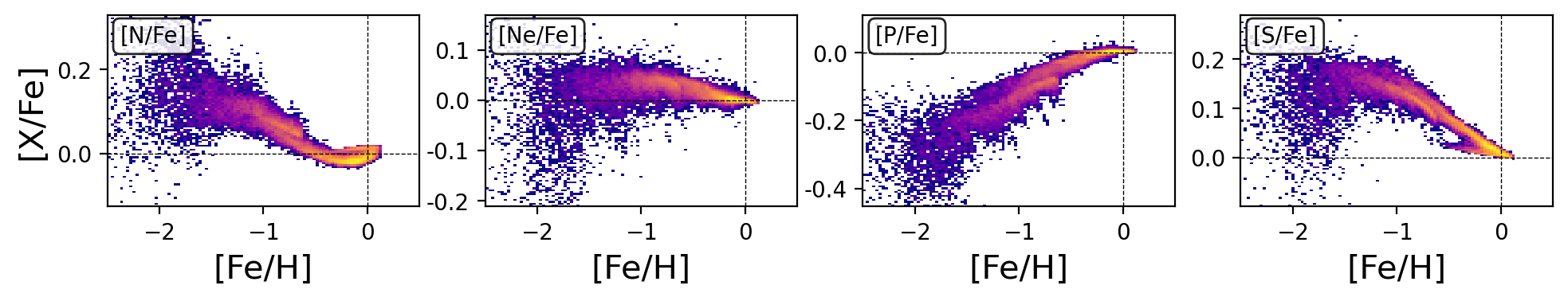}
    \caption{
    \textbf{Same as Fig.~\ref{fig:FeH_XFe}, but for the elements N, Ne, P, and S that were not measured by GALAH.}
    }
    \label{fig:appendix_feh_xfe_nneps}
\end{figure*}

\begin{figure*}
	\includegraphics[width=\textwidth]{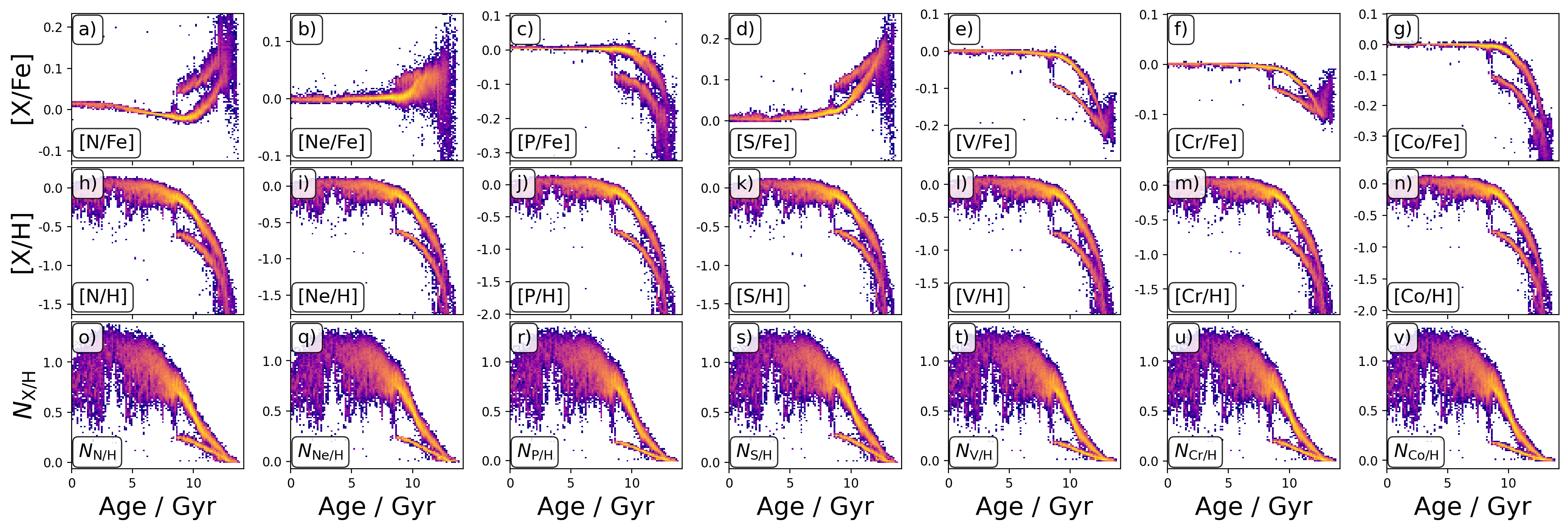}
    \caption{
    \textbf{Same as Fig.~\ref{fig:age_xfe_xh_nxh}, but for the elements N, Ne, P, S, V, Cr, and Co.}
    }
    \label{fig:appendix_xfe_xh_nx}
\end{figure*}

We append abundance-abundance figures (Fig.~\ref{fig:appendix_feh_xfe_nneps}) and age-abundance figures (Fig.~\ref{fig:appendix_xfe_xh_nx}) that were not shown in the main manuscript, mainly because the shown elements were either not measured in the GALAH Survey and would thus not provide further insight for the analyses of observations.

Furthermore, we append two figures: Firstly, Tab.~\ref{tab:tabular_separation_r12} of mean values and standard deviations of {[X/H]} in different age bins for the three age-abundance sequences of the Milky Way analogue and secondly, Tab.~\ref{tab:tabular_separation_r12_noise} of the separation significance $r_{12}$ between Sequence 1 and 2 in {[X/H]} for different elements and simulated observational noise across different age bins.

\begin{figure*}
	\includegraphics[width=\textwidth]{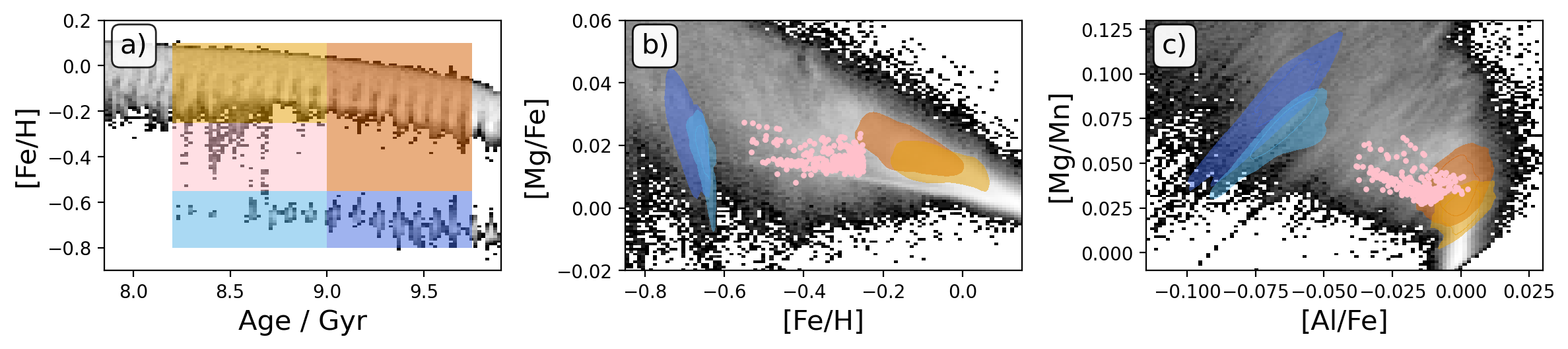}
    \caption{
    \textbf{Same as Fig.~\ref{fig:tracing_amr_regions_footprint} but for the whole Milky Way analogue.}
    }
    \label{fig:tracing_amr_regions_whole}
\end{figure*}

In addition, we append Fig.~\ref{fig:tracing_amr_regions_whole}, which extends the observations of Fig.~\ref{fig:tracing_amr_regions_footprint} to the whole Milky Way analogue.

\begin{table*}
    \centering
    \caption{\textbf{Mean values and standard deviations of {[X/H]} in different age bins for the three age-abundance sequences of the Milky Way analogue as identified in Eqs.~\ref{eq:sequence1}-\ref{eq:sequence3} as well as the separation significance $r_{12}$ between Sequence 1 and 2 as defined in Eq.~\ref{eq:r_value}.} Values are calculated for the observable footprint without observational noise if more than 10 particles are within a bin. We do not show Ne and P, as they are commonly not measured by stellar surveys.}
    \tiny
    \begin{tabular}{cccccccccccc}
\hline\hline
Element & Quantity  & 8.5-9.0 & 9.0-9.5 & 9.5-10.0 & 10.0-10.5 & 10.5-11.0 & 11.0-11.5 & 11.5-12.0 & 12.0-12.5 & 12.5-13.0 & 13.0-13.5 \\
\hline
[C/H] & $\mu_1 \pm \sigma_1$  & $-0.09 \pm 0.04$  & $-0.13 \pm 0.04$  & $-0.20 \pm 0.04$  & $-0.31 \pm 0.05$  & $-0.43 \pm 0.06$  & $-0.55 \pm 0.07$  & $-0.72 \pm 0.07$  & $-0.90 \pm 0.11$  & $-1.17 \pm 0.13$  & $-1.97 \pm 0.36$  \\
 & $\mu_2 \pm \sigma_2$  & $-0.54 \pm 0.02$  & $-0.57 \pm 0.02$  & $-0.61 \pm 0.02$  & $-0.69 \pm 0.03$  & $-0.78 \pm 0.04$  & $-0.89 \pm 0.05$  & $-1.03 \pm 0.09$  & $-1.31 \pm 0.15$  & $-1.66 \pm 0.24$  & $-2.10 \pm 0.21$  \\
 & $\mu_3 \pm \sigma_3$  & -  & -  & -  & -  & -  & $-1.24 \pm 0.17$  & $-1.57 \pm 0.10$  & $-1.72 \pm 0.17$  & -  & -  \\
  & $r_{12}$  & 10.5 & 9.9 & 8.9 & 6.6 & 4.8 & 4.0 & 2.8 & 2.1 & 1.8 & 0.3 \\
\hline
[N/H] & $\mu_1 \pm \sigma_1$  & $-0.13 \pm 0.04$  & $-0.19 \pm 0.04$  & $-0.26 \pm 0.04$  & $-0.37 \pm 0.05$  & $-0.49 \pm 0.06$  & $-0.60 \pm 0.06$  & $-0.74 \pm 0.07$  & $-0.89 \pm 0.09$  & $-1.14 \pm 0.11$  & $-1.76 \pm 0.31$  \\
 & $\mu_2 \pm \sigma_2$  & $-0.60 \pm 0.02$  & $-0.63 \pm 0.03$  & $-0.66 \pm 0.03$  & $-0.73 \pm 0.04$  & $-0.80 \pm 0.04$  & $-0.90 \pm 0.05$  & $-1.01 \pm 0.08$  & $-1.23 \pm 0.14$  & $-1.55 \pm 0.22$  & $-1.89 \pm 0.23$  \\
 & $\mu_3 \pm \sigma_3$  & -  & -  & -  & -  & -  & $-1.25 \pm 0.15$  & $-1.49 \pm 0.10$  & $-1.66 \pm 0.15$  & -  & -  \\
  & $r_{12}$  & 10.3 & 9.3 & 8.2 & 6.0 & 4.5 & 3.8 & 2.5 & 2.0 & 1.7 & 0.3 \\
\hline
[O/H] & $\mu_1 \pm \sigma_1$  & $-0.09 \pm 0.04$  & $-0.13 \pm 0.04$  & $-0.19 \pm 0.04$  & $-0.29 \pm 0.04$  & $-0.41 \pm 0.05$  & $-0.50 \pm 0.06$  & $-0.64 \pm 0.07$  & $-0.79 \pm 0.10$  & $-1.02 \pm 0.13$  & $-1.75 \pm 0.33$  \\
 & $\mu_2 \pm \sigma_2$  & $-0.57 \pm 0.02$  & $-0.59 \pm 0.02$  & $-0.62 \pm 0.02$  & $-0.68 \pm 0.02$  & $-0.76 \pm 0.03$  & $-0.85 \pm 0.04$  & $-0.97 \pm 0.07$  & $-1.21 \pm 0.16$  & $-1.50 \pm 0.23$  & $-1.89 \pm 0.16$  \\
 & $\mu_3 \pm \sigma_3$  & -  & -  & -  & -  & -  & $-1.24 \pm 0.20$  & $-1.55 \pm 0.11$  & $-1.70 \pm 0.16$  & -  & -  \\
  & $r_{12}$  & 11.3 & 11.4 & 10.4 & 7.8 & 5.5 & 4.7 & 3.4 & 2.2 & 1.8 & 0.4 \\
\hline
[Mg/H] & $\mu_1 \pm \sigma_1$  & $-0.10 \pm 0.04$  & $-0.15 \pm 0.04$  & $-0.22 \pm 0.04$  & $-0.33 \pm 0.05$  & $-0.46 \pm 0.06$  & $-0.56 \pm 0.06$  & $-0.72 \pm 0.07$  & $-0.89 \pm 0.11$  & $-1.14 \pm 0.14$  & $-1.93 \pm 0.37$  \\
 & $\mu_2 \pm \sigma_2$  & $-0.63 \pm 0.02$  & $-0.66 \pm 0.02$  & $-0.70 \pm 0.02$  & $-0.76 \pm 0.03$  & $-0.85 \pm 0.03$  & $-0.95 \pm 0.05$  & $-1.08 \pm 0.07$  & $-1.34 \pm 0.17$  & $-1.65 \pm 0.25$  & $-2.08 \pm 0.19$  \\
 & $\mu_3 \pm \sigma_3$  & -  & -  & -  & -  & -  & $-1.33 \pm 0.22$  & $-1.68 \pm 0.11$  & $-1.85 \pm 0.16$  & -  & -  \\
  & $r_{12}$  & 11.8 & 12.0 & 10.7 & 8.0 & 5.8 & 4.8 & 3.5 & 2.2 & 1.8 & 0.4 \\
\hline
[Al/H] & $\mu_1 \pm \sigma_1$  & $-0.11 \pm 0.04$  & $-0.17 \pm 0.04$  & $-0.25 \pm 0.04$  & $-0.37 \pm 0.05$  & $-0.51 \pm 0.06$  & $-0.63 \pm 0.07$  & $-0.81 \pm 0.08$  & $-1.01 \pm 0.12$  & $-1.30 \pm 0.15$  & $-2.18 \pm 0.39$  \\
 & $\mu_2 \pm \sigma_2$  & $-0.72 \pm 0.02$  & $-0.75 \pm 0.02$  & $-0.79 \pm 0.02$  & $-0.87 \pm 0.03$  & $-0.96 \pm 0.04$  & $-1.08 \pm 0.05$  & $-1.23 \pm 0.08$  & $-1.54 \pm 0.20$  & $-1.90 \pm 0.27$  & $-2.38 \pm 0.19$  \\
 & $\mu_3 \pm \sigma_3$  & -  & -  & -  & -  & -  & $-1.48 \pm 0.27$  & $-1.92 \pm 0.14$  & $-2.12 \pm 0.18$  & -  & -  \\
  & $r_{12}$  & 12.3 & 12.7 & 11.2 & 8.3 & 6.1 & 5.1 & 3.6 & 2.3 & 1.9 & 0.4 \\
\hline
[Si/H] & $\mu_1 \pm \sigma_1$  & $-0.10 \pm 0.04$  & $-0.15 \pm 0.04$  & $-0.22 \pm 0.04$  & $-0.33 \pm 0.05$  & $-0.45 \pm 0.06$  & $-0.56 \pm 0.06$  & $-0.71 \pm 0.07$  & $-0.88 \pm 0.10$  & $-1.13 \pm 0.13$  & $-1.85 \pm 0.33$  \\
 & $\mu_2 \pm \sigma_2$  & $-0.62 \pm 0.02$  & $-0.64 \pm 0.02$  & $-0.68 \pm 0.02$  & $-0.75 \pm 0.03$  & $-0.83 \pm 0.04$  & $-0.93 \pm 0.05$  & $-1.06 \pm 0.08$  & $-1.31 \pm 0.16$  & $-1.61 \pm 0.23$  & $-2.02 \pm 0.17$  \\
 & $\mu_3 \pm \sigma_3$  & -  & -  & -  & -  & -  & $-1.31 \pm 0.21$  & $-1.64 \pm 0.11$  & $-1.81 \pm 0.16$  & -  & -  \\
  & $r_{12}$  & 11.5 & 11.3 & 10.2 & 7.6 & 5.5 & 4.6 & 3.3 & 2.2 & 1.8 & 0.5 \\
\hline
[S/H] & $\mu_1 \pm \sigma_1$  & $-0.09 \pm 0.04$  & $-0.13 \pm 0.04$  & $-0.20 \pm 0.04$  & $-0.30 \pm 0.04$  & $-0.42 \pm 0.06$  & $-0.51 \pm 0.06$  & $-0.65 \pm 0.07$  & $-0.80 \pm 0.10$  & $-1.04 \pm 0.13$  & $-1.78 \pm 0.34$  \\
 & $\mu_2 \pm \sigma_2$  & $-0.57 \pm 0.02$  & $-0.59 \pm 0.02$  & $-0.62 \pm 0.02$  & $-0.68 \pm 0.03$  & $-0.76 \pm 0.03$  & $-0.86 \pm 0.05$  & $-0.98 \pm 0.08$  & $-1.23 \pm 0.17$  & $-1.54 \pm 0.24$  & $-1.96 \pm 0.17$  \\
 & $\mu_3 \pm \sigma_3$  & -  & -  & -  & -  & -  & $-1.25 \pm 0.20$  & $-1.57 \pm 0.10$  & $-1.74 \pm 0.16$  & -  & -  \\
  & $r_{12}$  & 11.1 & 10.8 & 9.8 & 7.4 & 5.3 & 4.4 & 3.1 & 2.2 & 1.8 & 0.5 \\
\hline
[V/H] & $\mu_1 \pm \sigma_1$  & $-0.12 \pm 0.05$  & $-0.18 \pm 0.04$  & $-0.27 \pm 0.04$  & $-0.40 \pm 0.06$  & $-0.55 \pm 0.07$  & $-0.69 \pm 0.07$  & $-0.88 \pm 0.08$  & $-1.10 \pm 0.12$  & $-1.40 \pm 0.13$  & $-2.08 \pm 0.31$  \\
 & $\mu_2 \pm \sigma_2$  & $-0.74 \pm 0.02$  & $-0.78 \pm 0.03$  & $-0.83 \pm 0.03$  & $-0.92 \pm 0.04$  & $-1.01 \pm 0.04$  & $-1.14 \pm 0.06$  & $-1.29 \pm 0.09$  & $-1.55 \pm 0.16$  & $-1.89 \pm 0.23$  & $-2.25 \pm 0.19$  \\
 & $\mu_3 \pm \sigma_3$  & -  & -  & -  & -  & -  & $-1.49 \pm 0.23$  & $-1.86 \pm 0.12$  & $-2.05 \pm 0.15$  & -  & -  \\
  & $r_{12}$  & 12.2 & 11.9 & 10.5 & 7.7 & 5.8 & 4.8 & 3.3 & 2.3 & 1.9 & 0.5 \\
\hline
[Cr/H] & $\mu_1 \pm \sigma_1$  & $-0.12 \pm 0.04$  & $-0.18 \pm 0.04$  & $-0.26 \pm 0.04$  & $-0.39 \pm 0.05$  & $-0.53 \pm 0.06$  & $-0.66 \pm 0.07$  & $-0.84 \pm 0.08$  & $-1.03 \pm 0.11$  & $-1.32 \pm 0.12$  & $-1.96 \pm 0.30$  \\
 & $\mu_2 \pm \sigma_2$  & $-0.70 \pm 0.02$  & $-0.74 \pm 0.03$  & $-0.78 \pm 0.03$  & $-0.86 \pm 0.04$  & $-0.95 \pm 0.04$  & $-1.07 \pm 0.06$  & $-1.20 \pm 0.09$  & $-1.45 \pm 0.16$  & $-1.77 \pm 0.22$  & $-2.13 \pm 0.18$  \\
 & $\mu_3 \pm \sigma_3$  & -  & -  & -  & -  & -  & $-1.42 \pm 0.20$  & $-1.74 \pm 0.11$  & $-1.93 \pm 0.15$  & -  & -  \\
  & $r_{12}$  & 11.8 & 11.2 & 10.0 & 7.3 & 5.5 & 4.6 & 3.1 & 2.2 & 1.8 & 0.5 \\
\hline
[Mn/H] & $\mu_1 \pm \sigma_1$  & $-0.13 \pm 0.04$  & $-0.18 \pm 0.04$  & $-0.27 \pm 0.04$  & $-0.40 \pm 0.06$  & $-0.54 \pm 0.07$  & $-0.67 \pm 0.07$  & $-0.86 \pm 0.08$  & $-1.06 \pm 0.11$  & $-1.35 \pm 0.12$  & $-1.96 \pm 0.30$  \\
 & $\mu_2 \pm \sigma_2$  & $-0.70 \pm 0.02$  & $-0.74 \pm 0.03$  & $-0.79 \pm 0.03$  & $-0.87 \pm 0.04$  & $-0.96 \pm 0.04$  & $-1.09 \pm 0.06$  & $-1.22 \pm 0.09$  & $-1.46 \pm 0.15$  & $-1.79 \pm 0.22$  & $-2.12 \pm 0.20$  \\
 & $\mu_3 \pm \sigma_3$  & -  & -  & -  & -  & -  & $-1.42 \pm 0.19$  & $-1.74 \pm 0.11$  & $-1.92 \pm 0.15$  & -  & -  \\
  & $r_{12}$  & 11.5 & 10.7 & 9.5 & 7.0 & 5.2 & 4.4 & 3.0 & 2.1 & 1.7 & 0.4 \\
\hline
[Fe/H] & $\mu_1 \pm \sigma_1$  & $-0.11 \pm 0.04$  & $-0.17 \pm 0.04$  & $-0.24 \pm 0.04$  & $-0.36 \pm 0.05$  & $-0.50 \pm 0.06$  & $-0.61 \pm 0.07$  & $-0.77 \pm 0.07$  & $-0.95 \pm 0.10$  & $-1.22 \pm 0.12$  & $-1.89 \pm 0.31$  \\
 & $\mu_2 \pm \sigma_2$  & $-0.65 \pm 0.02$  & $-0.68 \pm 0.03$  & $-0.72 \pm 0.03$  & $-0.80 \pm 0.03$  & $-0.88 \pm 0.04$  & $-0.99 \pm 0.05$  & $-1.12 \pm 0.08$  & $-1.37 \pm 0.16$  & $-1.69 \pm 0.23$  & $-2.06 \pm 0.19$  \\
 & $\mu_3 \pm \sigma_3$  & -  & -  & -  & -  & -  & $-1.36 \pm 0.20$  & $-1.67 \pm 0.10$  & $-1.86 \pm 0.15$  & -  & -  \\
  & $r_{12}$  & 11.4 & 10.8 & 9.7 & 7.2 & 5.3 & 4.4 & 3.1 & 2.2 & 1.8 & 0.5 \\
\hline
[Co/H] & $\mu_1 \pm \sigma_1$  & $-0.12 \pm 0.05$  & $-0.17 \pm 0.04$  & $-0.26 \pm 0.05$  & $-0.39 \pm 0.06$  & $-0.55 \pm 0.07$  & $-0.69 \pm 0.08$  & $-0.89 \pm 0.09$  & $-1.11 \pm 0.13$  & $-1.43 \pm 0.14$  & $-2.23 \pm 0.36$  \\
 & $\mu_2 \pm \sigma_2$  & $-0.76 \pm 0.02$  & $-0.80 \pm 0.03$  & $-0.85 \pm 0.03$  & $-0.94 \pm 0.04$  & $-1.04 \pm 0.04$  & $-1.18 \pm 0.06$  & $-1.33 \pm 0.09$  & $-1.63 \pm 0.18$  & $-1.98 \pm 0.25$  & $-2.42 \pm 0.20$  \\
 & $\mu_3 \pm \sigma_3$  & -  & -  & -  & -  & -  & $-1.54 \pm 0.26$  & $-1.97 \pm 0.14$  & $-2.17 \pm 0.17$  & -  & -  \\
  & $r_{12}$  & 12.5 & 12.6 & 11.1 & 8.2 & 6.1 & 5.0 & 3.6 & 2.3 & 1.9 & 0.5 \\
\hline
[Ba/H] & $\mu_1 \pm \sigma_1$  & $-0.10 \pm 0.05$  & $-0.17 \pm 0.04$  & $-0.26 \pm 0.05$  & $-0.40 \pm 0.07$  & $-0.58 \pm 0.09$  & $-0.83 \pm 0.12$  & $-1.18 \pm 0.13$  & $-1.64 \pm 0.23$  & $-2.24 \pm 0.17$  & $-3.12 \pm 0.35$  \\
 & $\mu_2 \pm \sigma_2$  & $-0.59 \pm 0.03$  & $-0.70 \pm 0.05$  & $-0.78 \pm 0.05$  & $-0.95 \pm 0.07$  & $-1.14 \pm 0.09$  & $-1.40 \pm 0.11$  & $-1.71 \pm 0.17$  & $-2.20 \pm 0.19$  & $-2.69 \pm 0.24$  & $-3.18 \pm 0.28$  \\
 & $\mu_3 \pm \sigma_3$  & -  & -  & -  & -  & -  & $-1.71 \pm 0.28$  & $-2.28 \pm 0.17$  & $-2.46 \pm 0.19$  & -  & -  \\
  & $r_{12}$  & 9.1 & 8.0 & 7.4 & 5.6 & 4.3 & 3.5 & 2.4 & 1.9 & 1.5 & 0.1 \\
\hline
\hline
\end{tabular}

    \label{tab:tabular_separation_r12}
\end{table*}

\begin{table*}
    \centering
    \caption{\textbf{Separation significance $r_{12}$ between Sequence 1 and 2 in {[X/H]} for different elements (first column) and simulated observational noise (second column) across different age bins (third column onwards, in Gyr).} Values are calculated for the observable footprint according to Eq.~\ref{eq:r_value}. For reference, we also add the median observational noise of GALAH DR3 in the first column for each element.}
    \begin{tabular}{cccccccccccc}
\hline\hline
Element & $\sigma_\mathrm{obs}$  & 8.5-9.0 & 9.0-9.5 & 9.5-10.0 & 10.0-10.5 & 10.5-11.0 & 11.0-11.5 & 11.5-12.0 & 12.0-12.5 & 12.5-13.0 & 13.0-13.5 \\
\hline
[C/H] & 0.05 & 5.4 & 5.3 & 4.9 & 4.2 & 3.4 & 3.1 & 2.4 & 2.0 & 1.7 & 0.3 \\
$\sigma_\mathrm{GALAH}$ & 0.10 & 3.0 & 3.0 & 2.8 & 2.5 & 2.2 & 2.1 & 1.7 & 1.7 & 1.6 & 0.3 \\
0.12 & 0.15 & 2.1 & 2.0 & 1.9 & 1.7 & 1.5 & 1.5 & 1.3 & 1.4 & 1.4 & 0.3 \\
\hline
[O/H] & 0.05 & 5.9 & 5.7 & 5.3 & 4.5 & 3.7 & 3.4 & 2.8 & 2.1 & 1.7 & 0.4 \\
$\sigma_\mathrm{GALAH}$ & 0.10 & 3.3 & 3.1 & 2.9 & 2.6 & 2.3 & 2.2 & 1.9 & 1.8 & 1.6 & 0.4 \\
0.13 & 0.15 & 2.2 & 2.1 & 2.0 & 1.8 & 1.6 & 1.6 & 1.4 & 1.5 & 1.4 & 0.3 \\
\hline
[Mg/H] & 0.05 & 6.4 & 6.2 & 5.7 & 4.9 & 4.0 & 3.6 & 2.9 & 2.1 & 1.8 & 0.4 \\
$\sigma_\mathrm{GALAH}$ & 0.10 & 3.6 & 3.5 & 3.2 & 2.9 & 2.5 & 2.4 & 2.1 & 1.8 & 1.6 & 0.4 \\
0.08 & 0.15 & 2.5 & 2.4 & 2.2 & 2.0 & 1.7 & 1.7 & 1.5 & 1.5 & 1.4 & 0.3 \\
\hline
[Al/H] & 0.05 & 7.0 & 6.9 & 6.3 & 5.4 & 4.4 & 4.0 & 3.1 & 2.2 & 1.9 & 0.4 \\
$\sigma_\mathrm{GALAH}$ & 0.10 & 4.0 & 3.9 & 3.6 & 3.2 & 2.8 & 2.7 & 2.3 & 1.9 & 1.7 & 0.4 \\
0.07 & 0.15 & 2.8 & 2.7 & 2.5 & 2.3 & 2.0 & 2.0 & 1.7 & 1.7 & 1.6 & 0.4 \\
\hline
[Si/H] & 0.05 & 6.2 & 6.0 & 5.5 & 4.7 & 3.8 & 3.5 & 2.7 & 2.1 & 1.8 & 0.4 \\
$\sigma_\mathrm{GALAH}$ & 0.10 & 3.5 & 3.4 & 3.1 & 2.8 & 2.4 & 2.3 & 2.0 & 1.8 & 1.6 & 0.4 \\
0.06 & 0.15 & 2.4 & 2.3 & 2.1 & 1.9 & 1.7 & 1.6 & 1.5 & 1.5 & 1.4 & 0.4 \\
\hline
[V/H] & 0.05 & 7.1 & 6.9 & 6.4 & 5.3 & 4.3 & 3.9 & 2.9 & 2.2 & 1.8 & 0.5 \\
$\sigma_\mathrm{GALAH}$ & 0.10 & 4.1 & 4.0 & 3.7 & 3.3 & 2.8 & 2.7 & 2.2 & 1.9 & 1.6 & 0.4 \\
0.13 & 0.15 & 2.8 & 2.8 & 2.6 & 2.3 & 2.0 & 2.0 & 1.7 & 1.6 & 1.4 & 0.4 \\
\hline
[Cr/H] & 0.05 & 6.7 & 6.5 & 5.9 & 5.0 & 4.0 & 3.6 & 2.7 & 2.1 & 1.7 & 0.5 \\
$\sigma_\mathrm{GALAH}$ & 0.10 & 3.9 & 3.7 & 3.5 & 3.1 & 2.6 & 2.5 & 2.0 & 1.8 & 1.6 & 0.4 \\
0.08 & 0.15 & 2.7 & 2.6 & 2.4 & 2.1 & 1.9 & 1.8 & 1.5 & 1.5 & 1.4 & 0.4 \\
\hline
[Mn/H] & 0.05 & 6.7 & 6.4 & 5.8 & 4.8 & 3.9 & 3.5 & 2.6 & 2.0 & 1.7 & 0.4 \\
$\sigma_\mathrm{GALAH}$ & 0.10 & 3.8 & 3.7 & 3.4 & 3.0 & 2.6 & 2.4 & 1.9 & 1.7 & 1.5 & 0.4 \\
0.08 & 0.15 & 2.6 & 2.6 & 2.4 & 2.1 & 1.8 & 1.8 & 1.5 & 1.4 & 1.3 & 0.4 \\
\hline
[Fe/H] & 0.05 & 6.3 & 6.1 & 5.6 & 4.7 & 3.8 & 3.4 & 2.6 & 2.0 & 1.8 & 0.5 \\
$\sigma_\mathrm{GALAH}$ & 0.10 & 3.6 & 3.5 & 3.2 & 2.8 & 2.4 & 2.3 & 1.9 & 1.7 & 1.6 & 0.4 \\
0.07 & 0.15 & 2.5 & 2.4 & 2.2 & 2.0 & 1.7 & 1.7 & 1.4 & 1.4 & 1.4 & 0.4 \\
\hline
[Co/H] & 0.05 & 7.3 & 7.2 & 6.7 & 5.6 & 4.6 & 4.1 & 3.1 & 2.2 & 1.8 & 0.5 \\
$\sigma_\mathrm{GALAH}$ & 0.10 & 4.3 & 4.2 & 3.9 & 3.5 & 3.0 & 2.9 & 2.4 & 2.0 & 1.7 & 0.4 \\
0.08 & 0.15 & 2.9 & 2.9 & 2.7 & 2.4 & 2.2 & 2.1 & 1.8 & 1.7 & 1.5 & 0.4 \\
\hline
[Ba/H] & 0.05 & 5.5 & 5.5 & 5.2 & 4.5 & 3.8 & 3.2 & 2.3 & 1.8 & 1.5 & 0.1 \\
$\sigma_\mathrm{GALAH}$ & 0.10 & 3.2 & 3.4 & 3.3 & 3.2 & 2.9 & 2.7 & 2.0 & 1.7 & 1.4 & 0.1 \\
0.09 & 0.15 & 2.2 & 2.4 & 2.3 & 2.3 & 2.3 & 2.1 & 1.7 & 1.5 & 1.2 & 0.1 \\
\hline
\hline
\end{tabular}

    \label{tab:tabular_separation_r12_noise}
\end{table*}

\begin{figure*}
	\includegraphics[width=\textwidth]{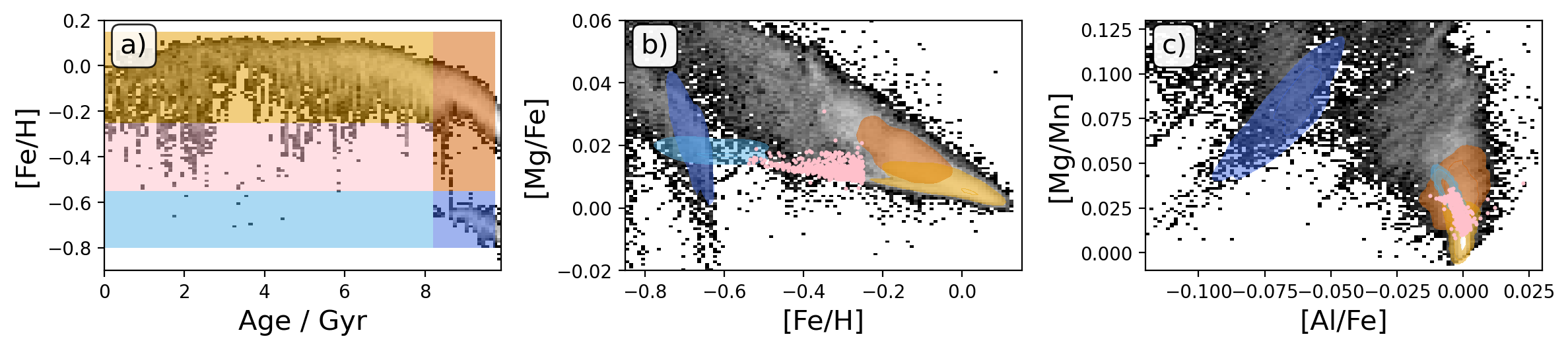}
    \caption{
    \textbf{Same as Fig.~\ref{fig:tracing_amr_regions_footprint} but also tracing the youngest stars of the Milky Way analogue.} Note that the traced groups are separated at the end of the merger ($8.2\,\mathrm{Gyr}$), rather than before it ($9.0\,\mathrm{Gyr}$).
    }
    \label{fig:tracing_amr_regions_disk}
\end{figure*}

\bsp	
\label{lastpage}
\end{document}